\documentclass[12pt]{iopart}

\usepackage{graphicx}

\usepackage{color}

\usepackage[normalem]{ulem}

\usepackage{cite}

\usepackage[pscoord]{eso-pic}

\newcommand{\placetextbox}[3]{
  \setbox0=\hbox{#3}
  \AddToShipoutPictureFG*{
    \put(\LenToUnit{#1\paperwidth},\LenToUnit{#2\paperheight}){\vtop{{\null}\makebox[0pt][c]{#3}}}
  }
}

\begin{document}

\placetextbox{0.5}{0.98}{\large\textcolor{red}{Version of the preprint of the paper published in}}%
\placetextbox{0.5}{0.96}{\large\textcolor{red}{Superconductor Science \& Technology {\bf 32}, (2019) 045009}}%
\placetextbox{0.5}{0.94}{\large\textcolor{red}{with an extended introduction and more references.}}%

\title[The conductivity and the magnetization around $T_c$ in OPT Y-123...]{The conductivity and the magnetization around $T_c$ in optimally-doped YBa$_2$Cu$_3$O$_{7-\delta}$ revisited: quantitative analysis in terms of fluctuating superconducting pairs}

\author{R I Rey\footnote{Present address: Clinical Neurosciences Research Laboratory, Clinical University Hospital, Health Research Institute of Santiago de Compostela (IDIS),  Universidade de Santiago de Compostela, Spain}, C Carballeira, J M Doval\footnote{Present address: Clinical Neurosciences Research Laboratory, Clinical University Hospital, Health Research Institute of Santiago de Compostela (IDIS), Universidade de Santiago de Compostela, Spain}, J Mosqueira, M V Ramallo, A Ramos-\'{A}lvarez\footnote{Present address: ICFO-Institut de Ciencies Fot\`oniques, The Barcelona Institute of Science and Technology, 08860 Castelldefels (Barcelona), Spain}, D S\'o\~nora, J A Veira, J C Verde, F Vidal}

\address{Quantum Materials and Photonics Research Group, Particle Physics Department, University of Santiago de Compostela, ES-15782 Santiago de Compostela, Spain}


\vspace{10pt}
\begin{indented}
\item[]July 2018
\end{indented}

\begin{abstract}
We first present detailed measurements of the rounding behavior around the superconducting transition temperature, $T_c$, of the in-plane electrical conductivity, magnetoconductivity and magnetization, including the low and moderate magnetic field regimes, in a high-quality single crystal and a thin film of the prototypical optimally-doped YBa$_2$Cu$_3$O$_{7-\delta}$ (OPT Y-123), in which the inhomogeneity effects are minimized. Then, it is presented a comparison of these experimental data with the phenomenological Ginzburg-Landau (GL) approach that takes into account the unavoidable contribution of the fluctuating pairs, the only theoretical scenario allowing at present to analyze these roundings at the quantitative level. These analyses demonstrate that the measured rounding effects around $T_c$ may be explained quantitative and consistently in terms of the GL scenario, even up to the rounding onset temperatures if the quantum localization, associated with the shrinkage of the superconducting wavefunction, is taken into account. The implications of our results on the pseudogap physics of optimally-doped cuprates  are also discussed.
\end{abstract}

\section{Introduction}

The phenomenological descriptions of the superconducting transition of the high-tem\-per\-a\-ture cuprate superconductors (HTSC) remain at present not well settled, despite their interest being considerably enhanced by the absence of a well-established {understanding of the} microscopic pairing mechanism for these superconductors.\cite{ref1a,ref1b} A central still open issue concerning this phenomenology is the description of the rounding behavior shown around the superconducting critical temperature, $T_c$, by different observables. It was early proposed that at least in optimally-doped cuprates (OPT HTSC) these roundings are due to the presence of conventional superconducting fluctuations, i.e., to the presence of fluctuating superconducting pairs created by the unavoidable thermal agitation energy \cite{ref2,ref3a,ref3b,ref4,ref5,ref6}. In fact, in a wide temperature interval, above but not too close to $T_c$ (above the so-called Ginzburg-Levanyuk temperature) and well below the rounding onset temperature, these effects in  OPT HTSC were well described in terms of the phenomenological mean-field Gaussian-Ginzburg-Landau (GGL) approach for the  superconducting fluctuations of both the amplitude and the phase of the order parameter in multilayered superconductors \cite{ref2,ref3a,ref3b,ref4,ref5,ref6}. This conclusion was supported, e.g., by measurements of the magnetization rounding above $T_c$  (the so-called precursor diamagnetism) \cite{ref5,ref6,ref7a,ref7b} and of the in-plane electrical resistivity rounding (the so-called paraconductivity) \cite{ref5,ref6,ref8,ref9,ref10a,ref10b,ref10c,ref10d,ref10e,ref10f,ref10g,ref10h,ref10i,ref10j,ref10k,ref10l}. These two observables are not only crucial to characterize the superconducting transition but, in addition, in HTSC they present an excellent ratio between the extent of the rounding effects and the corresponding background or bare contributions (orders of magnitude bigger than, for instance, in the heat capacity) \cite{ref2,ref6}.

In applying the  mean field GGL approach to the roundings observed above the superconducting transition in HTSC, it was used as critical temperature, $T_c$, the experimental transition temperature at which both the Meissner and the resistivity transitions occur. Moreover, as first stressed in Ref.~\cite{ref11}, this approach predicts that in the mean-field-like region where it is applicable, the fluctuations effects on both observables, the paraconductivity and the precursor diamagnetism, will have the same dependence on the reduced temperature, $\varepsilon\equiv\ln(T/T_c)$. The confirmation of this prediction by measurements in  different OPT HTSC (a result that will be extended in our present work to the high reduced temperature region, $\varepsilon>0.1$) not only provided further support to the GGL scenario but also was an early indication of the absence of the so-called indirect fluctuation effects on both observables \cite{ref11}. This conclusion concerns in particular the Maki-Thompson contribution to the in-plane paraconductivity \cite{ref2,ref3a,ref3b,ref6,ref11}, in agreement with the enhanced pair-breaking effects expected for the non-$^1s_0$ wave pairing in the HTSC \cite{ref12a,ref12b}. The GL scenario was further supported by early scaling analyses of the electrical conductivity and the diamagnetism measured around $T_c(H)$ in the so-called critical region \cite{ref13a,ref13b,ref13c,ref13d,ref13e}. This GL scenario was also found compatible with the existence close to the zero-field $T_c$ of a full-critical region where the mean-field-like approaches are no longer applicable. In this full-critical region the superconducting fluctuations have been described in terms of the XY model \cite{ref8,ref9,ref14,ref15a,ref15b}.

The GL scenario for multilayered superconductors was afterward applied to some HTSC with doping levels different from the optimal, although the conclusions could be deeply affected by the unavoidable presence of appreciable chemical inhomogeneities.\cite{ref16,ref17} Importantly, the  applicability of the mean field GGL approaches was then extended to the high reduced temperature region by empirically introducing a so-called total-energy cutoff,\cite{ref18,ref19,ref20} that takes into account the unavoidable Heisenberg localization energy associated with the shrinkage of the superconducting wavefunction when the temperature increases above the superconducting transition.\cite{ref21} However, the corresponding {early} calculations predict{ed} a different temperature behavior for the paraconductivity and the precursor diamagnetism in the high reduced temperature region, an erroneous conclusion that is corrected now in our present work (a result that in turn is, as we will see, confirmed at a quantitative level by the high quality experimental data presented here). Other aspects of these comparisons with the GGL approaches may be seen in Refs.~\cite{ref22} and \cite{ref23}.

Mainly in the last ten years, however, different research groups, including some of the most influential, have proposed other physical mechanisms, different from the presence of conventional superconducting fluctuations, to account for the entire or a part of the measured rounding effects above $T_c$ in all the HTSC, including in those compounds  where (as is the case of the optimally-doped YBa$_2$Cu$_3$O$_{7-\delta}$, OPT Y-123), the rounding onsets are located relatively close to $T_c$, below twice the measured superconducting critical temperature.\cite{ref1a,ref1b,ref24,ref25,ref26,ref27,ref28,ref29,ref30a,ref30b,ref30c,ref30d,ref30e,ref30f,ref30m,ref30o,ref30p,ref30l,ref30g,ref30n,ref30q,ref30r,ref30i,ref30j,taillefer} In some cases, these proposals were fueled by the observation in different HTSC, even in OPT Y-123, of seemingly intrinsic anomalous roundings around $T_c$, non-describable in terms of the GL approaches summarized above. These anomalous roundings concern not only the in-plane magnetization \cite{ref24,ref25,ref26,ref27} and the $dc$ in-plane electrical conductivity \cite{ref28,ref29}, but also other observables, as the Nernst effect \cite{ref27,ref30a,ref30b,ref30c,ref30d,ref30e,ref30f,ref30m,ref30o,ref30p,ref30l,ref30g,ref30n,ref30q,ref30r,ref30i,ref30j,taillefer}. 
Although some of these anomalies could in fact be attributed to $T_c$ inhomogeneities, associated with the presence of chemical disorder with long characteristics lengths (larger than the superconducting coherence length amplitude), unavoidable in non-stoichiometric compounds \cite{ref30k,ref31,ref32,ref33,ref34,ref35,ref36}, most of the works published since then on that issue still attribute the measured rounding effects above $T_c$ to the opening of a pseudogap in the normal state.\cite{ref24,ref25,ref26,ref27,ref28,ref29,ref30a,ref30b,ref30c,ref30d,ref30e,ref30f,ref30m,ref30o,ref30p,ref30l,ref30g,ref30n,ref30q,ref30r,ref30i,ref30j,taillefer,ref30k,ref31,ref32,ref33,ref34,ref35,ref36,ref37a,ref37b,ref37c,ref38,ref39,ref40,ref41,ref42,ref43,ref44,ref45,ref46,ref47,ref48,ref49,ref50,ref51,ref52} In this last scenario, the most popular mechanism is based on phase disordering.\cite{ref1a,ref1b,ref24,ref25,ref26,ref27,ref28,ref29,ref30a,ref30b,ref30c,ref30d,ref30e,ref30f,ref30m,ref30o,ref30p,ref30l,ref30g,ref30n,ref30q,ref30r,ref30i,ref30j,ref37a,ref37b,ref37c,ref38,ref39,ref40,ref41,ref42,ref43,ref44,ref45,ref46,ref47,ref48,ref49,ref50,ref51,ref52} In these so-called phase incoherent superconductivity approaches the true pairing temperature will be well above the measured $T_c$ (where the Meissner and the resistivity transitions occurs), in some cases up to the temperature at which the normal state pseudogap opens. Such a seemingly broad region of short-range phase stiffness above the measured $T_c$, and up to the pseudogap temperature, would support large vorticity, and the corresponding effects should be then much more important than those predicted by the conventional approaches based on the presence of fluctuating pairs above the mean-field critical temperature.\cite{ref1a,ref1b,ref24,ref37a,ref37b,ref37c,ref38,ref39,ref40,ref41,ref42,ref43,ref44,ref45,ref46,ref47,ref48,ref49,ref50,ref51,ref52}

The difficulties to discriminate between these very different theoretical scenarios proposed for the rounding effects around $T_c$ in HTSC are to some extent related to the unavoidable experimental uncertainties that arise when extracting the intrinsic roundings from the measurements. These uncertainties have in turn various, and often entangled, sources: First at all, those associated with the always somewhat arbitrary approximations needed to determine both the $T_c$ of each theoretical approach and the background or bare contributions, associated with the normal state behavior, to each observable, a difficulty that concerns the descriptions of any phase transition.\footnote{The uncertainties associated with the background or bare contributions affect all the comparisons with the theoretical approaches for any second order phase transition in any system, and they have their origin in the fact that these approaches calculate the fluctuation contributions separately from the non-critical contributions [see, {\it e.g.}, Ref.~\cite{ref57}]. So, these uncertainties are to some extent intrinsic to any experimental background estimation, included those obtained by {trying to quench} the superconductivity by high reduced magnetic fields.} Secondly, the presence of $T_c$-inhomogeneities, mainly those associated with chemical disorder, these last being in some cases intrinsic-like, due to the unavoidable random distribution of dopant atoms.\cite{ref31,ref32,ref33,ref34,ref35,ref36} In addition, the layered nature of the HTSC and the complexity of their chemistry increases the relevance of these $T_c$-inhomogeneities, whose influence may be particularly dramatic in compounds with doping different from the optimal.\cite{ref30k,ref31,ref32,ref33,ref34,ref35,ref36}
Indeed, in the case of the theoretical approaches associated with the opening of a pseudogap in the normal state, these difficulties are further enhanced by the scarcity until now of quantitative theoretical proposals for the observed effects, in contrast with the GL approaches for the fluctuations of superconducting pairs.\cite{ref2,ref3a,ref3b,ref4,ref5,ref6,ref7a,ref7b,ref13a,ref13b,ref13c,ref13d,ref13e} Moreover, in the temperature region where these superconducting fluctuations, thermodynamic-like, are always present (typically below 2$T_c$), the corresponding rounding effects should be taken into account when considering other possible effects. 

The difficulties commented before, some of them unavoidable, affect even the discrimination between the different versions of the GL scenarios proposed until now. In fact, in spite of earlier detailed measurements and analyses,\cite{ref3a,ref3b,ref4,ref5,ref6,ref7a,ref7b,ref8,ref9,ref10a,ref10b,ref10c,ref10d,ref10e,ref10f,ref10g,ref10h,ref10i,ref10j,ref10k,ref10l,ref11,ref12a,ref12b,ref13a,ref13b,ref13c,ref13d,ref13e,ref14,ref15a,ref15b,ref16,ref17,ref18,ref19,ref20,ref21,ref22,ref23} aspects so central within these approaches as the possible presence of indirect fluctuation effects, the multilayering influence and the estimations of the corresponding effective periodicity lengths, the role of the chemical disorder and of the corresponding $T_c$-inhomogeneities, the behavior at high reduced temperatures and magnetic fields, or the physical origin of the corresponding cutoffs,  are still controversial.\cite{ref29,ref30a,ref30b,ref30c,ref30d,ref30e,ref30f,ref30m,ref30o,ref30p,ref30l,ref30g,ref30n,ref30q,ref30r,ref30i,ref30j,taillefer,ref30k,ref31,ref32,ref33,ref34,ref35,ref36,ref58,ref59,ref60,ref61,ref62,ref63,ref64a,ref64c,ref64b}

In this paper, the experimental uncertainties commented above are strongly mitigated by measuring the rounding behavior of the in-plane electrical conductivity, the magnetoconductivity and the magnetization around $T_c$ in high quality OPT Y-123 samples. These measurements extend up to 300~K in temperature and  up to 9~T in applied magnetic field amplitudes, that when analyzed on the grounds of the GL approaches cover both their low magnetic field limit (Schmidt regime) and moderate fields (or Prange) regime. In addition to their relevance to probe the superconducting transition, these observables were chosen because, as noted before, in HTSC they present an excellent ratio between the amplitudes of the rounding effects and the corresponding background or bare contributions associated with the normal state behavior. In turn, as also stressed before, the choice of the OPT Y-123 superconductor was first motivated by the high chemical and structural quality of the samples that may be prepared, what minimizes the $T_c$-inhomogeneities. In addition, the normal-state resistivity and the magnetic susceptibility are linear in an extended temperature region, allowing a simple and reliable extraction of the background. In fact, these characteristics make OPT Y-123 the prototypical compound to analyze the rounding effects in HTSC, and in the last twenty years these effects were extensively measured on it and analyzed in terms of superconducting fluctuations.\cite{ref5,ref6,ref7a,ref7b,ref8,ref9,ref10a,ref10b,ref10c,ref10d,ref10e,ref10f,ref10g,ref10h,ref10i,ref10j,ref10k,ref10l,ref11,ref12a,ref12b,ref13a,ref13b,ref13c,ref13d,ref13e,ref14,ref15a,ref15b,ref16,ref17,ref18,ref19,ref20,ref21,ref24,ref25,ref26,ref27,ref28,ref29,ref30a,ref30b,ref30c,ref30d,ref30e,ref30f,ref30m,ref30o,ref30p,ref30l,ref30g,ref30n,ref30q,ref30r,ref30i,ref30j,taillefer,ref30k,ref58,ref59,ref61} 
However, as stressed before, our present experimental results on so central  observables are performed in high quality samples and cover different  magnetic field regimes and reduced temperatures, including the rounding onset temperature itself. They are, therefore, unique and particularly useful to discriminate between the different theoretical scenarios proposed until now. The presentation of these high-precision experimental data and the corresponding model-independent conclusions, that include a direct determination of the roundings' onset temperatures, is the first aim of this paper. 

Our second central aim is to present quantitative and consistent analyses of these experimental results in terms of the phenomenological Ginzburg-Landau (GL) approaches that take into account the fluctuating pairs contribution, by introducing thermal fluctuations of both the amplitude and the phase of the superconducting order parameter. Two main reasons justify the choice of these comparisons: The presence of fluctuating superconducting pairs is unavoidable around any superconducting transition and, in addition, until now the GL approaches provide the only scenario in which the current state of the theoretical calculations allows {fully} quantitative and simultaneous confrontations with explicit expressions for these rounding effects on so central observables. The comparisons summarized here include the extended (with a total-energy cutoff) Gaussian-Ginzburg-Landau (GGL) approach for multilayered superconductors above but not too close to $T_c$ (above the Ginzburg-Levanyuk reduced temperature) and, consistently, the GL scaling in the lowest-Landau-level (GL-LLL) approach for 3D superconductors in the critical region around $T_c(H)$. The total energy cutoff takes into account the quantum localization energy, associated with the shrinkage of the superconducting wavefunction, and allows us to extend  the comparisons with the GL scenario up to the rounding onset itself. Two first conclusions for this GL scenario are going to be obtained by just comparing to each other the measured observables: the absence of indirect fluctuation effects on the in-plane paraconductivity and of non-local electrodynamic effects on the precursor diamagnetism, including the high reduced-temperature region. These  analyses are going to demonstrate at a quantitative level, and for the first time up to the rounding onset temperatures, that the measured roundings around $T_c$ in the prototypical OPT Y-123 may be explained consistently in terms of the phenomenological GL scenario, based on the presence of fluctuating pairs. They will provide also precise information on the corresponding parameters arising in this GL scenario, including the effective interlayer periodicity length, the in-plane superconducting coherence length amplitude, the rounding onset temperature and the corresponding cutoff, and the relaxation time of the superconducting pairs relative to the BCS value. 

The importance of the results presented here, as well as the unique role played by the prototypical OPT Y-123, are further enhanced by the fact that they provide some compelling thermodynamic constraints for the different scenarios being proposed to understand the so-called pseudogap of the HTSC, at present a debated issue and whose implications concern some of the proposals for the superconducting pairing origin.\cite{ref1a,ref1b,ref24,ref25,ref26,ref27,ref28,ref29,ref30a,ref30b,ref30c,ref30d,ref30e, ref30f,ref30m,ref30o,ref30p,ref30l,ref30g,ref30n,ref30q,ref30r,ref30i,ref30j,taillefer,Shekhter13,Badoux16} For instance, our results indicate that in OPT Y-123 the superconducting fluctuations are conventional and, therefore,  at odds with pseudogap scenarios that assume, for OPT doping, unconventionaly large roundings due to anomalous phase-dominated superconducting fluctuations. In this paper we will also briefly comment on some of the scenarios for the pseudogap that are instead compatible with our present results, including, e.g., those in which the pseudogap temperature, $T^*$, falls at optimal doping below the superconducting dome. Some other  proposed scenarios for the measured rounding effects, such as the possibility  of charge density wave effects \cite{Blanco14,Wu15,Badoux16}, are also going to be briefly discussed. 

The paper is organized as follows. In Sect. 2, we summarize the main aspects of the samples fabrication and details of the experimental procedures to perform the measurements of the in-plane electrical conductivity, the magnetoconductivity and the magnetization. In Sect. 3, we provide information on the procedures used to characterize the rounding effects around $T_c$ and to extract the model and background-independent results on the paraconductivity, magnetoconductivity and diamagnetism. The analyses at a quantitative level of these experimental results in terms of the GL approaches extended to high reduced temperatures is summarized in Sect. 4. We also propose a schematic $H-T$ phase diagram for the superconducting fluctuation scenario around $T_c(H)$ in OPT Y-123.
In Section 5, it is briefly commented on other possible scenarios for the observed rounding effects, in particular those associated with the presence of charge density waves and, mainly, with the pseudogap. In Sect. 6, these different results are summarized, stressing that they do not support that in OPT Y-123 the rounding effects above but near $T_c$ are, total or partially,  due to the opening of a pseudogap in the normal state, as it is being proposed by different relevant research groups. On the contrary, our results, in particular in the high reduced temperature region, further support that these roundings are originated by the presence of fluctuating superconducting pairs, as in conventional low $T_c$ superconductors and, then, that in OPT Y-123 the pseudogap will open well below $T_c$. It will be also stressed in that Section that, nevertheless, these conclusions not only cannot be extrapolated to the underdoped and overdoped HTSC but, in fact, that the in-plane resistivity roundings observed well above 2$T_c$ in underdoped HTSC cannot be attributed to superconducting fluctuations. Finally, in the Appendix, it is presented the extension to high reduced-temperatures of the GGL approach for the fluctuation conductivity and the precursor diamagnetism in 2D and layered superconductors.

\section{Fabrication of the samples and details of the measurements}

\subsection{Thin {film} and measurements of {the} electrical conductivity and magnetoconductivity}

To determine the electrical conductivity and the magnetoconductivity roundings around $T_c$ we have grown OPT Y-123 thin films over (100) SrTiO$_3$ substrates by high-pressure $dc$ sputtering. Details of the growth and characterization of these films are similar to those described elsewhere.\cite{ref65} Microbridges were patterned  on those films by photolithography and wet chemical etching into a four-probe configuration suitable for the electrical measurements. Au pads were then deposited onto the current and voltage terminals. The final contact resistance achieved was less than 0.1~$\Omega$/contact. The in-plane resistivity ($\rho_{ab}$) was measured in presence of magnetic fields up to 9 T perpendicular to the $ab$-layers with a Quantum Design's physical property measurement system (PPMS) by using an excitation current of $\sim50$~$\mu$A at 23 Hz. The uncertainties in the geometry and dimensions of the microbridges lead to an uncertainty in the $\rho_{ab}(T)$ amplitude below 20\%.

%
%
\begin{figure}[t]
\begin{center}
\includegraphics[scale=0.7]{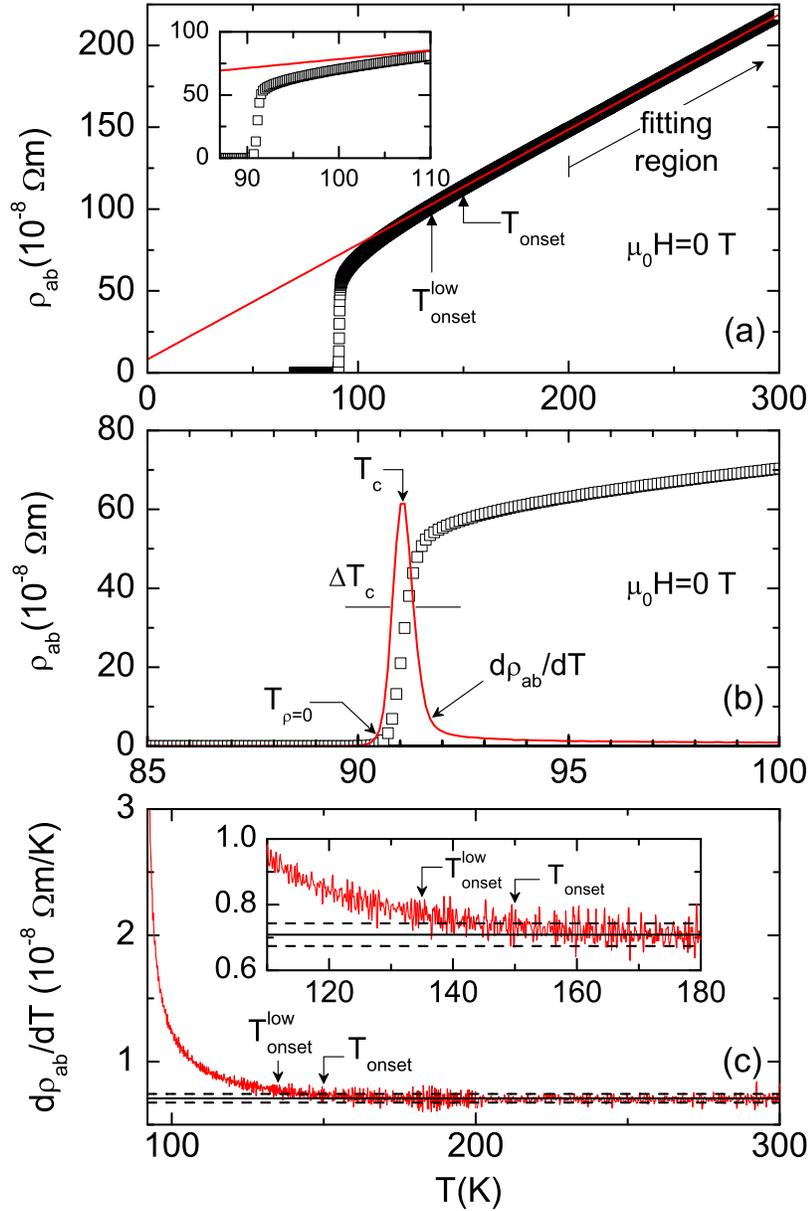}
\caption{(a) The in-plane resistivity versus temperature under zero applied magnetic field measured in the microbridge described in the main text. The solid line is the background contribution. Inset: Detail around $T_c$. (b) $\rho_{ab}(T)$ curve around $T_c$ together with its temperature derivative (in arbitrary units). (c) Overview of $d\rho_{ab}(T)/dT$ showing the onset temperature of the resistivity rounding ($T_{\rm onset}$) and its lower limit ($T_{\rm onset}^{\rm low}$). Inset: Detail around $T_{\rm onset}$.}
\label{f1}
\end{center}
\end{figure}

An example of the $\rho_{ab}(T)$ curves obtained in the microbridge studied in our present work is shown in Fig.~\ref{f1}(a). This microbridge was 2.1~mm long, 90~nm thick, and 40~$\mu$m wide. A detail of the resistive transition is presented in Fig.~\ref{f1}(b), where the solid line is d$\rho_{ab}(T)/$d$T$. The superconducting transition temperature $T_c$, and the resistive transition width, $\Delta T_c$, are obtained as the temperature location of the maximum of  the derivative peak and, respectively, its width at half maximum. For this microbridge, $T_c=91.10$~K  and $\Delta T_c=0.55$~K. Complementarily, the temperature at which the resistance becomes unobservable is denoted $T_{\rho=0}$, and is of the order of 90.5~K. These results are summarized in Table I.

The linear temperature dependence of the normal-state resistivity well above $T_c$, between 150 K and 300 K, further confirms the high quality of our Y-123 film. This is illustrated in Figs.~\ref{f1}(a) and (c). Such a linear behaviour was already observed in high-quality OPT Y-123 single-crystals and epitaxial thin films by different groups.\cite{ref5,ref6,ref8,ref9,ref10a,ref10b,ref10c,ref10d,ref10e,ref10f,ref10g,ref10h,ref10i,ref10j,ref10k,ref10l,ref11,ref13a,ref13b,ref13c,ref13d,ref13e,ref14,ref15a,ref15b,ref16,ref20,ref66,ref67} In addition, our microbridges present a normal-state resistivity ratio $\rho_{ab}(300{\rm K})/\rho_{ab}(100{\rm K})$ around 3.1, as expected for OPT Y-123 \cite{Wuyts96}. Also, the extrapolation to $T=0$~K of the normal state resistivity leads to very low residual values, as shown in Fig.~\ref{f1}(a). These results already suggest the absence of appreciable $T_c$-inhomogeneities in our microbridges. In fact, the possible presence of a small amount of non-superconducting domains uniformly distributed in the samples could just affect the paraconductivity amplitude, a relatively small effect that may be expected to remain well below the uncertainties associated with the microbridges dimensions.\cite{ref68}\footnote{Note that an incomplete superconducting volume fraction ($f<1$) would affect both the electrical resistivity and the magnetization, and then the extraction of the corresponding paraconductivity and precursor diamagnetism. In the case of the magnetization, independently of the spatial distribution of the non-superconducting domains, their presence will manifest only in the corresponding $\Delta\chi(T,H)$ amplitude (and then only the $H_{c2}(0)$ value resulting from the comparison with the GGL approach will be affected). The effects of the non-superconducting domains are more difficult to evaluate when extracting $\Delta\sigma_{ab}(\varepsilon)$. Assuming that there are non-superconducting domains uniformly distributed, probably the most usual case, the effective electrical conductivity may be approximated by using an effective medium approach\cite{ref68}
\begin{equation}
\int \frac{\sigma-\sigma_{\rm eff}}{\sigma+2\sigma_{\rm eff}}  Q(\sigma) {\rm d}\sigma=0,\nonumber
\end{equation}
where Q$(\sigma)$ is the conductivity volume distribution. This expression may be recast into
\begin{equation}
\sigma_{\rm eff}=\int \sigma Q(\sigma)  {\rm d}\sigma- \int \frac{(\sigma-\sigma_{\rm eff})^2}{\sigma+2\sigma_{\rm eff}}  Q(\sigma) {\rm d}\sigma.\nonumber
\end{equation}
Then, to the first order in $(\sigma-\sigma_{\rm eff})/\sigma_{\rm eff}$ the effective conductivity will be the volume-weighted average of the local conductivity. By assuming the presence of two types of domains, one with $\sigma_1=1/\rho_B+\Delta\sigma$ and a superconducting volume fraction $f$, and other non superconducting with $\sigma_2=1/\rho_B$ and a volume fraction $1-f$ it follows $\sigma_{\rm eff}= f\sigma_1+(1-f)\sigma_2$. Then the effective paraconductivity would be given by $\Delta\sigma_{\rm eff}\equiv\sigma_{\rm eff}-\sigma_B=f\Delta\sigma$ \textit{i.e.}, it is also modulated in amplitude by $f$ (a fact that could affect the $\tau_{\rm rel}$ value resulting from the comparison with the experimental data, see subsection~\ref{subsec-comparisonGGL}). Therefore, the quantity $-\Delta\chi(\varepsilon)_H/T\Delta\sigma_{ab}(\varepsilon)$ would be almost unaffected by a possible incomplete superconducting volume fraction. }

%
%
\begin{table}[t]
\begin{center}
\begin{tabular}{ccccccc}
\hline
  &  $T_c$ (K)    &   $\Delta T_c $(K)  &   $T_{\rm onset}$ (K) & $\Delta T_{\rm onset}$ (K) \\
\hline 
$\sigma_{ab}(T)$ & 91.1 &  0.6 &  150 &  15  \\
$\chi(T)_{H\rightarrow0}$ &  90.5      &  1    &    150  &  15    \\ 
\hline
\end{tabular}
\end{center}
\caption{The critical temperature, $T_c$, and the onset temperature of the rounding effects, $T_{\rm onset}$, together with the corresponding dispersions, $\Delta T_c$ and $\Delta T_{\rm onset}$, of the two optimally doped (OPT) Y-123 samples studied in this work. The data obtained from the resistivity measurements correspond to a thin film, whereas those obtained from the magnetization correspond to a single-crystal. For details see the main text.}
\end{table}

\begin{figure}[t]
\begin{center}
\includegraphics[scale=.7]{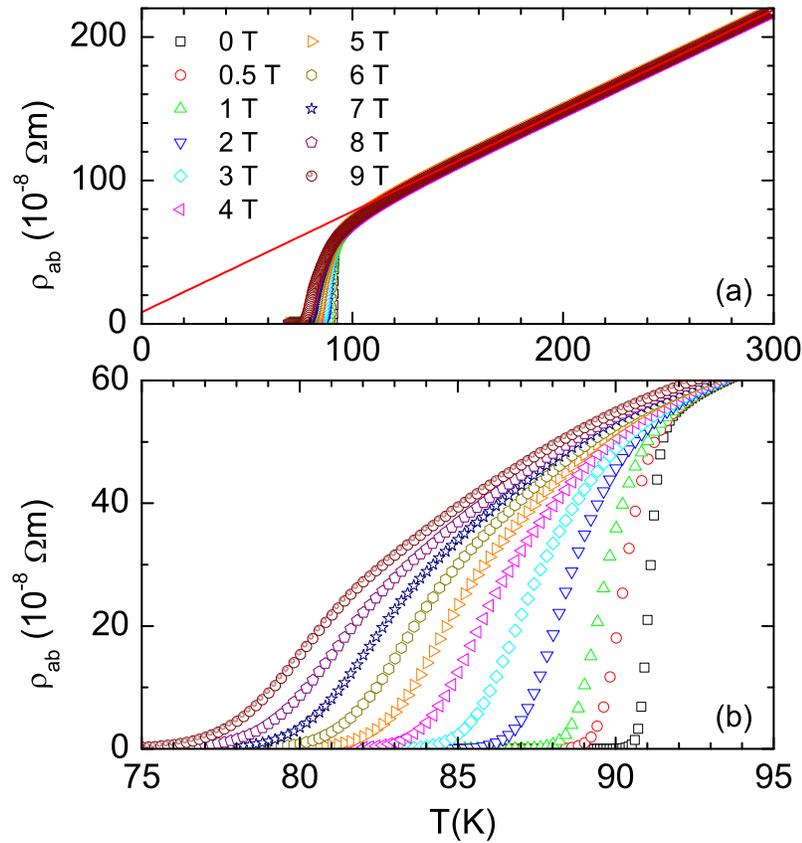}
\caption{(a) Temperature dependence of the in-plane resistivity measured under magnetic fields up to 9 T applied perpendicularly to the $ab$-layers. The solid line is the background contribution. (b) Detail around $T_c$. }
\label{f2}
\end{center}
\end{figure}

Some examples of the in-plane resistivity (for magnetic fields, $H$, applied perpendicularly to the superconducting $ab$-layers) are presented in Figs.~\ref{f2}(a) and (b). The data of Fig.~\ref{f2}(a) show that above 150~K the magnetoresistivity is inappreciable. This behaviour contrasts with the important magnetoresistivity observed near $T_c$, as illustrated by the results summarized in Fig.~\ref{f2}(b).

%
%

\subsection{Single-{crystal} and measurements of the magnetization}

To measure the magnetization with $H$ applied perpendicular to the superconducting $ab$-layers, $M(T,H)$, we have used a high-quality OPT Y-123 single-crystal prepared following a previously reported self-flux technique.\cite{ref9} It has a quite regular plate-like shape, with dimensions $2.3\times2.9$~mm$^2$ in the $ab$ plane and a thickness of 40~$\mu$m. The mass of this sample is 1.74~mg. The magnetization measurements were performed with a magnetometer based in the superconducting quantum interference (Quantum Design, model MPMS-XL). 

A first magnetic characterization of our crystal was obtained through the temperature dependence of the zero-field-cooling (ZFC) magnetic susceptibility ($\chi$) under a low magnetic field (0.25~mT) perpendicular to the $ab$ layers (Fig.~3(a)). From this curve one may obtain the diamagnetic transition temperature, $T_c$, as the temperature at which d$\chi/$d$T$ is maximum ($T_c=90.5$~K). Also, the diamagnetic transition width, $\Delta T_c$, is estimated as the width at half maximum of the d$\chi/$d$T$ peak, and for this example $\Delta T_c=1$~K. These results are also summarized in Table I. The differences between these $T_c$ and $\Delta T_c$ values and the ones corresponding to the microbridge studied in the previous subsection are well within the dispersion expected for different high-quality OPT Y-123 samples.\cite{ref6,ref7a,ref7b,ref8,ref9,ref10a,ref10b,ref10c,ref10d,ref10e,ref10f,ref10g,ref10h,ref10i,ref10j,ref10k,ref10l,ref11,ref13a,ref13b,ref13c,ref13d,ref13e,ref14,ref15a,ref15b,ref16,ref17,ref18,ref19,ref20,ref21}

The magnetic susceptibility rounding above $T_c$, that following the conventional denomination will already be called \textit{precursor diamagnetism}, is orders of magnitude smaller than in the Meissner region.\cite{ref2,ref6} Therefore, to measure these roundings one must apply magnetic fields orders of magnitude larger than those used in the ZFC measurements. Figure~3(b) provides an example of these measurements, performed under a field amplitude of 1~T. To assure a complete temperature stabilization, data are taken after waiting $\sim5$ minutes since the temperature was within 0.5\% the target temperature. In the case of the isothermal curves, it is waited around 5 minutes after setting the magnetic field in permanent mode. In both cases, the measurements were performed with the reciprocating sample option (RSO), that performs sinusoidal oscillations of the sample about the centre of the detection system, and improves the resolution with respect to the conventional $dc$ option. Each data point is the average of 4-6 individual measurements, each one consisting in 10 cycles at 1 Hz. The final resolution in magnetic moment was in the 10$^{-8}$ emu range.

Two aspects of these measurements deserve to be commented. First, as earlier stressed elsewhere,\cite{ref31} the presence in our {crystal} of a relatively small amount on non-superconducting domains with large characteristic lengths (much larger than the superconducting coherence length amplitude) will, independently of their spatial distribution, just affect the amplitude of the magnetization rounding observed above $T_c$, \textit{i.e.}, the precursor diamagnetism amplitude. Therefore, to simplify the analyses presented in the next sections (in particular when comparing with the paraconductivity) these possible effects will be neglected. When comparing with the GGL approaches, this approximation will in turn mainly affect the value of the amplitude of the upper critical field, but the corresponding uncertainties are expected to be comparable with those associated with the sample volume.  Note also that the linear temperature dependence of the normal-state magnetic susceptibility observed well above $T_c$ in the inset of Fig.~3(b) further confirms the high quality of our crystal. These last data correspond to measurements under an applied magnetic field of 1 T but, as shown by the data in Fig.~4, this linearity remains for field amplitudes up to 5~T.

%
%
\begin{figure}[t]
\begin{center}
\includegraphics[scale=.6]{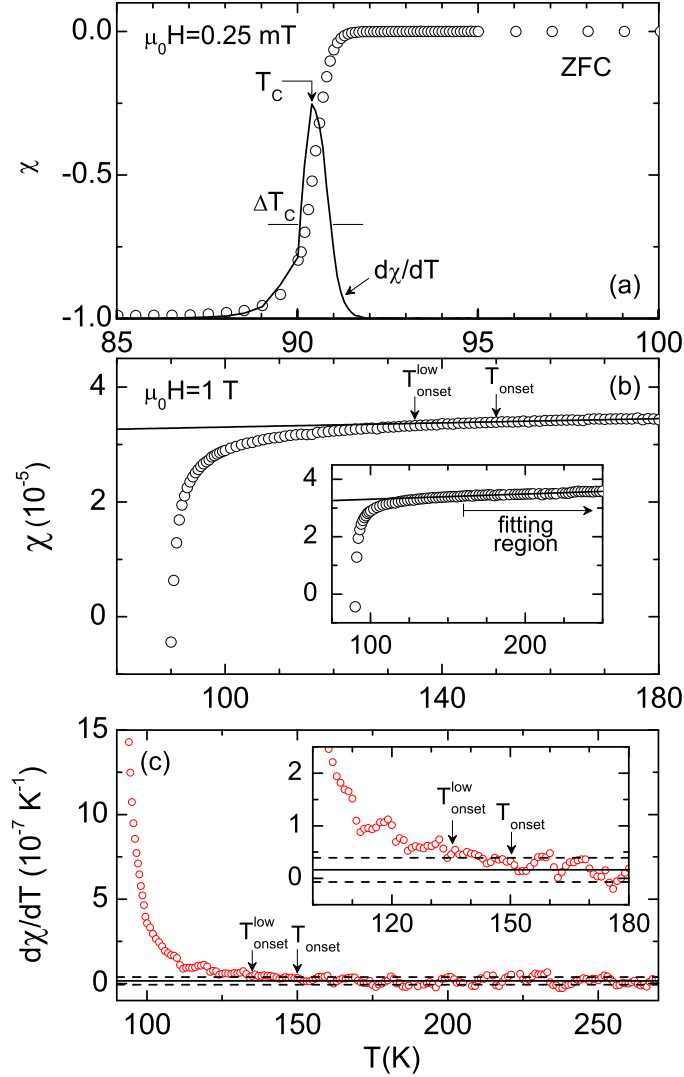}
\caption{(a) Detail around the superconducting transition of the magnetic susceptibility measured in the single crystal described in the main text. These measurements were performed by using a low amplitude magnetic field (0.25~mT) applied perpendicularly to the \textit{ab}-layers after zero-field-cooling (ZFC). The solid curve is $d\chi/dT$ in arbitrary units. The corresponding $T_c$, and the transition width, $\Delta T_c$, are also indicated. (b) An example of the magnetic susceptibility versus temperature measured with a 1~T magnetic field applied perpendicularly to the $ab$-layers. An overview up to room temperature is shown in the inset. The background contribution is represented as a solid line. As in the example for $\rho_{ab}(T)$ shown in Fig. 1, the background fitting region extends up to well above the onset of the fluctuation effects ($T_{\rm onset}$). It is also indicated here, as $T_{\rm onset}^{\rm low}$, the lower limit for this onset. (c) Temperature dependence of $d\chi/dT$ for $\mu_0H=1$~T, and the corresponding $T_{\rm onset}$ and $T_{\rm onset}^{\rm low}$. Inset: Detail around $T_{\rm onset}$.}
\label{f3}
\end{center}
\end{figure}

%
%
\begin{figure}[t]
\begin{center}
\includegraphics[scale=.7]{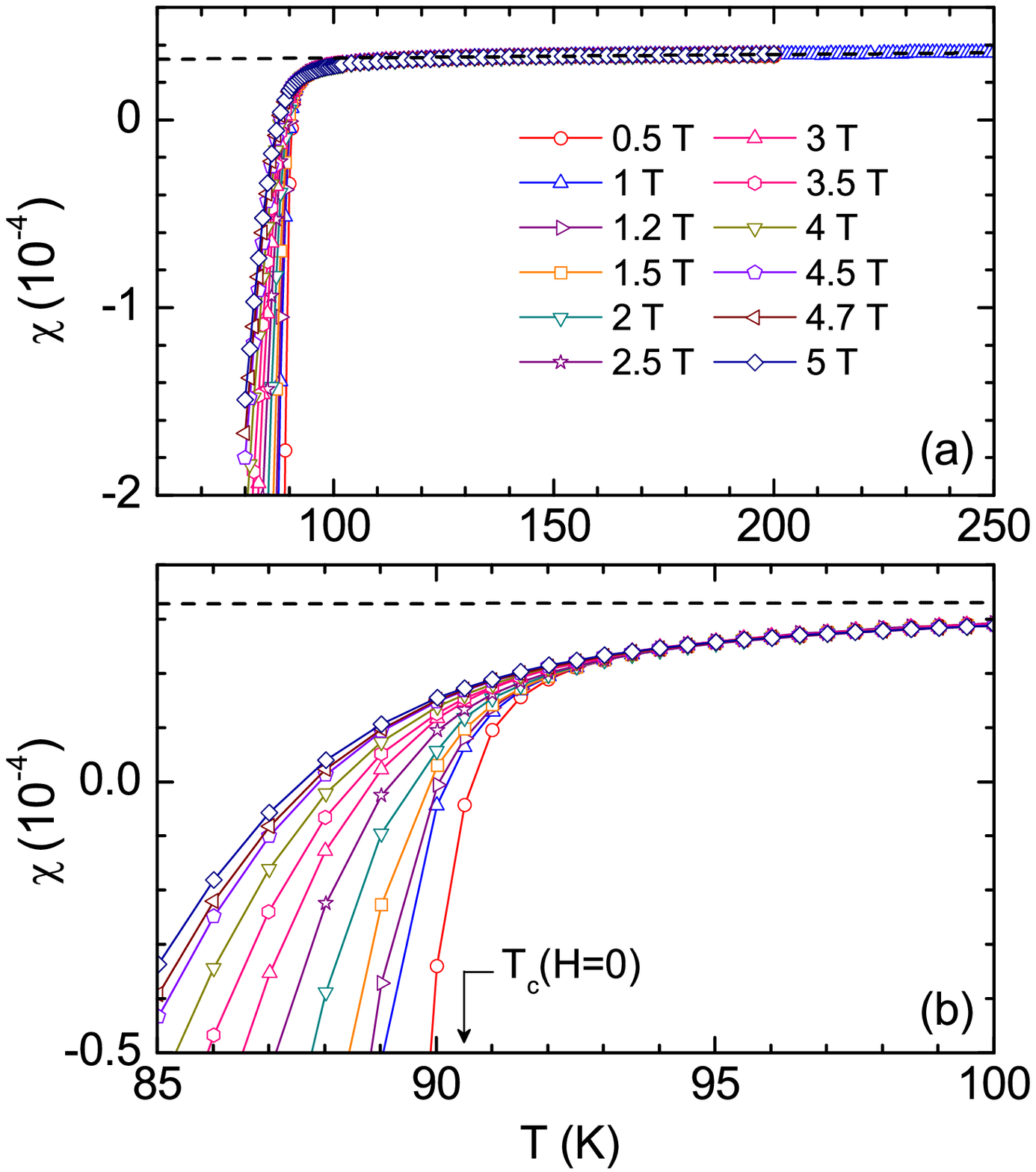}
\caption{(a) Temperature dependence of the magnetic susceptibility measured under magnetic fields up to 5 T applied perpendicularly to the $ab$-layers. (b) Detail around $T_c$. The dashed line is the background contribution. }
\label{f4}
\end{center}
\end{figure}

\section{{Characterization of the rounding effects:} Model- and background-independent results}

\subsection{Onset temperature of the {rounding effects}}

As shown in the precedent Section, well above $T_c$ and up to at least 250 K both the normal-state resistivity and the magnetic susceptibility depend linearly on temperature. For each observable, one may define the onset temperature of the rounding effects above $T_c$, denoted $T_{\rm onset}$, as the temperature below which {the measured behavior separates from the one that may be extrapolated from the high temperature region.} At the practical level, $T_{\rm onset}$ may be determined as the temperature at which either d$\rho_{ab}/$d$T$ or d$\chi/$d$T$ begin to rise above the extrapolated normal-state behavior beyond the noise level, as illustrated in Figs.~\ref{f1}(c) and \ref{f3}(c). We have also defined a lower bound for $T_{\rm onset}$, denoted $T_{\rm onset}^{\rm low}$, as the temperature below which practically all the $d\chi/dT$ or $d\rho/dT$ data fall outside the extrapolated (normal-state) noise band. Under this criteria, we have found that $T_{\rm onset}\approx150\pm15$~K for both the paraconductivity and the precursor diamagnetism, the latter for $\mu_0H=1$~T (a field small enough to still correspond to the so-called zero field limit, see next Section).

We have checked that the $T_{\rm onset}$ value determined from our resistivity measurements by using a background-independent procedure agrees, well within the corresponding experimental uncertainties, with those that may be inferred from the measurements of other research groups in the best available OPT Y-123 single-crystals and epitaxial films.\cite{ref8,ref9,ref10a,ref10b,ref10c,ref10d,ref10e,ref10f,ref10g,ref10h,ref10i,ref10j,ref10k,ref10l,ref11,ref13a,ref13b,ref13c,ref13d,ref13e,ref14,ref15a,ref15b} Similarly, we have checked that the $T_{\rm onset}$ we propose here from our magnetization measurements in the zero-field limit agrees with the results on the precursor diamagnetism onset that may be inferred from the $\chi(T)_H$ measurements of other authors in single-crystal and polycrystalline OPT Y-123 superconductors.\cite{ref6,ref7a,ref7b,ref19,ref20,ref21}

In Table~I we summarize the above results on $T_{\rm onset}$, together with those on $T_c$, obtained from the $\rho_{ab}(T)$ and $\chi(T)_H$ curves [in this last case under field amplitudes small enough so that $\chi(T)_H$ is field-independent, see Subsection 3.4]. Let us further stress here that these results are independent not only of a specific model for the superconducting fluctuations but also of a background substraction. This robustness enhances the importance of the agreement, well within the experimental uncertainties, of the $T_{\rm onset}$ values for these two observables. It is also remarkable the universality of the $T_{\rm onset}/T_c$ ratio, which takes a value similar to the one found in other OPT HTSC and in conventional low-$T_c$ superconductors.\cite{ref18,ref21,ref76}

\subsection{Paraconductivity}

\begin{figure}[t]
\begin{center}
\includegraphics[scale=.6]{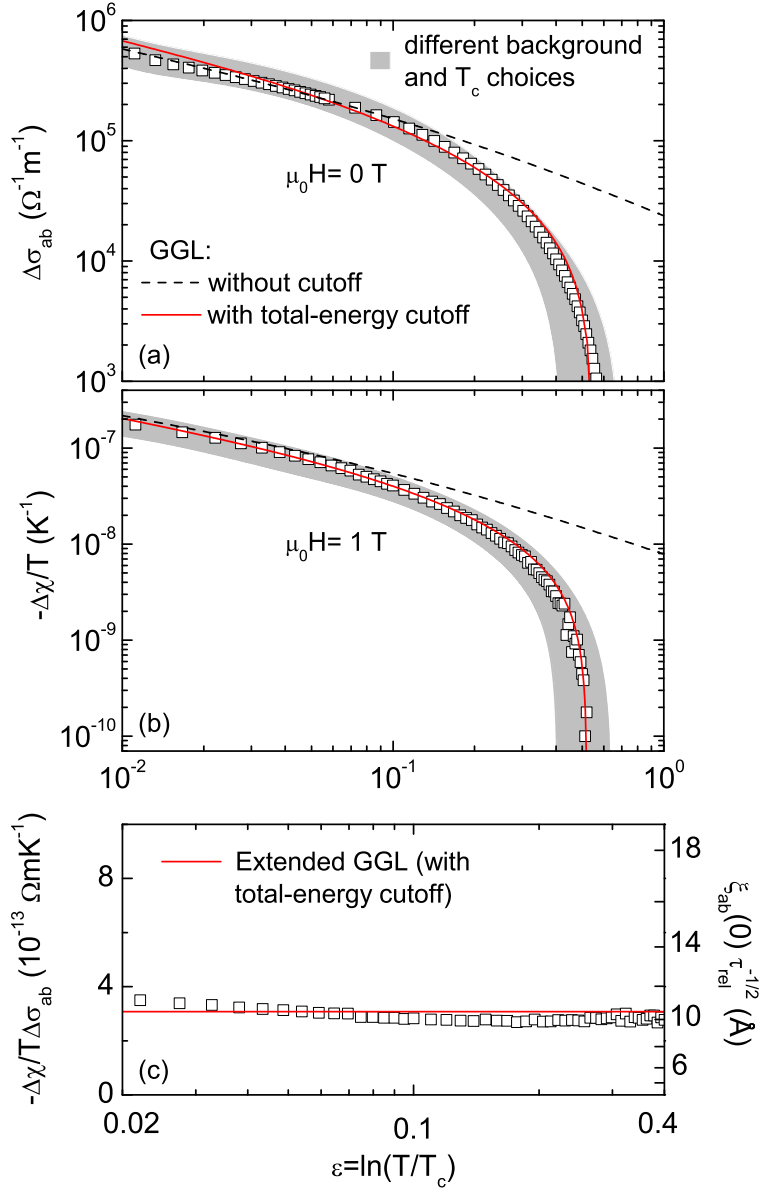}
\caption{(a) $\varepsilon$-dependence of $\Delta\sigma_{ab}$ extracted from the results shown in Fig.~1(a). The dashed line corresponds to the GGL approach in the zero field limit without cutoff [Eq.~\ref{eq11}] with $\tau_{\rm rel}= 1$ and $\xi_c(0)$= 0.13 nm. The solid line corresponds to Eq.~\ref{eq9} (which includes a total-energy cutoff) with $\tau_{\rm rel}= 1$, $\xi_c(0)$= 0.10 nm and $\varepsilon^c$= 0.55. (b) The $-\Delta\chi/T$ dependence on $\varepsilon$ for the OPT Y-123 single-crystal in the zero-magnetic-field. The dashed and solid lines are the GGL approaches for layered superconductors: without any cutoff [Eq.~\ref{eq12}, dashed line] and with a total-energy cutoff [Eq.~\ref{eq10}, solid line], using $\xi_c(0)$= 0.11 nm, $\xi_{ab}(0)$= 1.1 nm and $\varepsilon^c$= 0.52. (c) $\varepsilon$-dependence of $-\Delta\chi/T\Delta\sigma$ compared with Eq.~\ref{eq13} (solid line), with $\xi_{ab}^2(0)/\tau_{\rm rel}\approx 1.1 $ nm$^2$ as the only free parameter. The uncertainties associated with the background contribution and $T_c$ are represented by the shaded areas. }
\label{f5}
\end{center}
\end{figure}

{The rounding effect on the zero-field in-plane electrical conductivity, which following the conventional denomination will already be called \textit{paraconductivity}, is defined as\cite{ref2}}
\begin{eqnarray}
\Delta\sigma_{ab}(T)\equiv\sigma_{ab}(T)-\sigma_{abB}(T)=\frac{1}{\rho_{ab}(T)}-\frac{1}{\rho_{abB}(T)}.
\label{eq1}
\end{eqnarray}
Here $\rho_{abB}(T)$ is the background resistivity, estimated by linear extrapolation of the $\rho_{ab}(T)$ data above $T_{\rm onset}$, as shown in Fig.~1(a). In this figure, the background fitting region is $200-250$~K ({\it i.e.},  $2.2-2.7T_c$), thus avoiding the possible superlinear behavior associated with the CuO chains near room temperature \cite{Gagnon94}. The resulting $\Delta\sigma_{ab}(\varepsilon)$, where $\varepsilon\equiv\ln(T/T_c)$ is the reduced temperature, is shown in Fig.~\ref{f5}(a). 

To probe the robustness of these paraconductivity data, we have first checked that they are stable against changes of both the extension and the location of the background fitting region, if this region remains between $T_{\rm onset}$ and 300~K.  We have also checked that $\Delta\sigma_{ab}(\varepsilon)$ remains stable when  changing the critical temperature between $T_c$ and $T_c\pm\Delta T_c$. All these $\Delta\sigma_{ab}(\varepsilon)$ uncertainties correspond to the shadowed region in Fig. 5(a) (that around $T_{\rm onset}$, \textit{i.e.}, $\varepsilon \approx$ 0.55, covers the uncertainties estimated for $T_{\rm onset}$ in the precedent subsection).

The $\Delta\sigma_{ab}(\varepsilon)$ data of Fig. 5(a) agree at a quantitative level with most of the measurements published by different authors in OPT Y-123,\cite{ref6,ref8,ref9,ref10a,ref10b,ref10c,ref10d,ref10e,ref10f,ref10g,ref10h,ref10i,ref10j,ref10k,ref10l,ref18,ref28,ref29,ref58,ref59,ref60,ref61} although in general these previous results may be not so accurate in the high-$\varepsilon$ region.  Two aspects make our new data of particular interest: First, they already provide a further but robust experimental demonstration, also model-independent, of the sharp fall-off of $\Delta\sigma_{ab}(\varepsilon)$ when approaching $T_{\rm onset}$ from below.  Moreover, in contrast with other proposals,\cite{ref60,ref62} our present results confirm that such a fall-off does not follow a power-law in reduced temperature. Second, {these results may be confronted with the magnetoconductivity roundings} measured in the same film and with the precursor diamagnetism measured in a single-crystal. We will see in the last subsection the important, model-independent, information that those comparisons may provide. 

%
%

\subsection{{Magnetoconductivity rounding}}

{The magnetoconductivity rounding,} $\Delta\widetilde{\widetilde{\sigma}}_{ab}(\varepsilon,H)$, is defined as the change of the paraconductivity under a magnetic field, {\it i.e.},
\begin{eqnarray}
\Delta\widetilde{\widetilde{\sigma}}_{ab}(\varepsilon, H) &\equiv&\Delta\sigma(\varepsilon,H)-\Delta\sigma(\varepsilon,0)\nonumber\\
&=&\left[\sigma_{ab}(\varepsilon, H)-\sigma_{abB}(\varepsilon,H)\right]-\left[\sigma_{ab}(\varepsilon)-\sigma_{abB}(\varepsilon)\right].
\label{eq2}
\end{eqnarray}
The important point here is that in OPT Y-123, for magnetic fields below 10~T the normal-state magnetoconductivity may be neglected,\cite{ref5,ref6,ref89,ref89b,ref89c,ref29} i.e., $\sigma_{abB}(\varepsilon, H) \approx \sigma_{abB}(\varepsilon,0)$, as confirmed at a quantitative level by the results summarized in Fig.~2. Then $\Delta\widetilde{\widetilde{\sigma}}_{ab}(\varepsilon, H)$ may be approximated as the total magnetoconductivity, $\Delta\widetilde{\sigma}_{ab}(\varepsilon, H)$:   
\begin{eqnarray}
\Delta\widetilde{\widetilde{\sigma}}_{ab}(\varepsilon, H) \approx 
\Delta\widetilde{\sigma}_{ab}(\varepsilon, H) \equiv 
\sigma_{ab}(\varepsilon,H)-\sigma_{ab}(\varepsilon).
\label{eq3}
\end{eqnarray}
Therefore, $\Delta\widetilde{\widetilde{\sigma}}_{ab}(\varepsilon, H)$ may be estimated as the difference between two directly measurable observables.

Examples of the $\varepsilon$- and $H$-dependences of the paraconductivity, for the same microbridge as in Figs.~1 and 5(a), are summarized in Figs.~6(a) and (b). The $H$-dependence of the fluctuation-induced magnetoconductivity, $-\Delta\widetilde{\widetilde{\sigma}}_{ab}(\varepsilon,H)$, obtained from the same measurements by using Eq.~\ref{eq3}, is presented in Fig. 6(c). To better appreciate their general behavior, in this last case we have represented only the data points of three isotherms. These last results do not depend of any background choice and  they provide a further, model-independent, indication of the sharp decrease of the superconducting fluctuations above $\varepsilon\approx0.3$, in agreement with the paraconductivity under zero [Fig. 5(a)] and finite applied magnetic fields [Figs. 6(a) and 6(b)].

%
%
\begin{figure}[t]
\begin{center}
\includegraphics[scale=0.7]{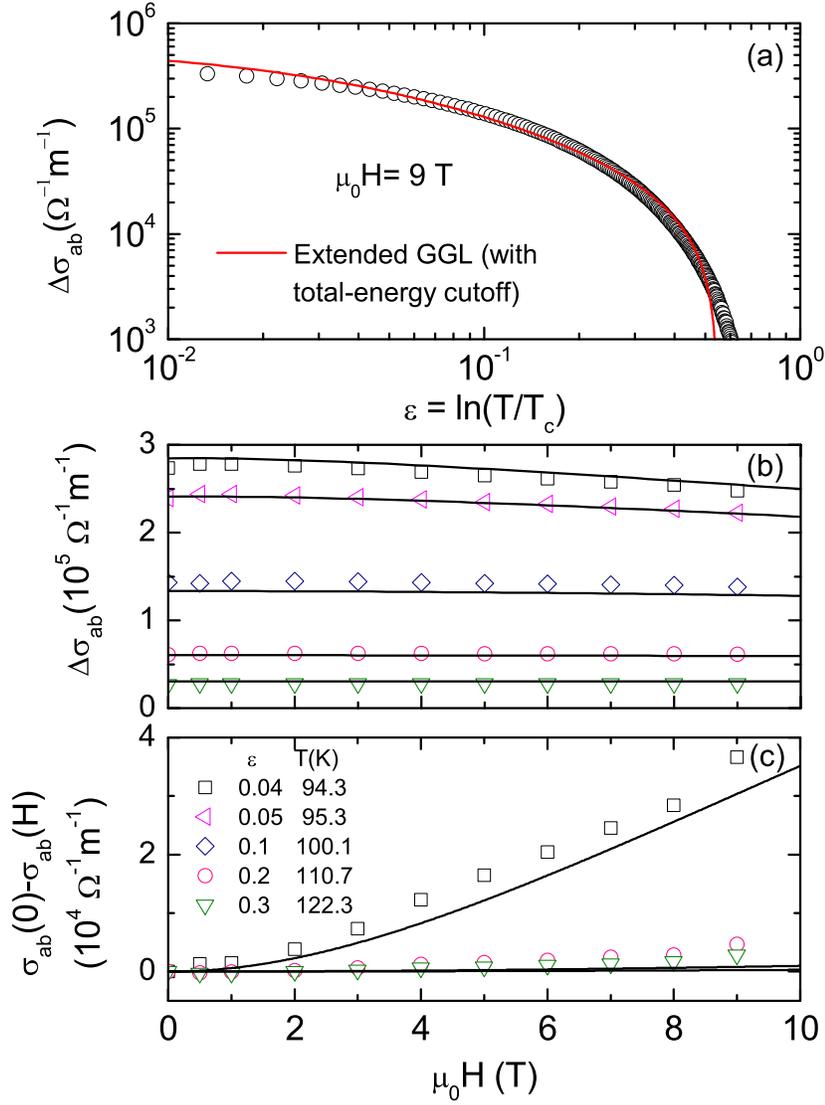}
\caption{(a) $\varepsilon$-dependence of the in-plane paraconductivity measured under an external magnetic field of 9 T. The line corresponds to the best fit of Eq.~\ref{eq7}  with the same $\xi_c(0)$ and $\varepsilon^c$ values that those obtained before from the paraconductivity for $H=0$ (Fig.~\ref{f5}(a)), and with $\xi_{ab}(0)$ as free parameter. This leads to $\xi_{ab}(0)$= 1.2 nm. (b) and (c) Magnetic field dependence of the paraconductivity and the magnetoconductivity for different temperatures above $T_c$. The lines correspond to $\Delta\sigma_{ab}(\varepsilon,H)$ and, respectively $\Delta\sigma_{ab}(\varepsilon)-\Delta\sigma_{ab}(\varepsilon,H)$, evaluated by using  Eq.~\ref{eq7} with the same $\xi_{ab}(0)$ and $\xi_c(0)$ values as before. The agreement is reasonable taking into account that $\sigma_{ab}(\varepsilon,0)-\sigma_{ab}(\varepsilon,H)$ is typically one order of magnitude smaller than $\Delta\sigma$, and that it is affected by the individual uncertainties of both $\sigma_{ab}(\varepsilon,0)$ and $\sigma_{ab}(\varepsilon,H)$.}  
\label{f6}
\end{center}
\end{figure}

\subsection{Precursor diamagnetism}

The {rounding} effects on the magnetic susceptibility for $H \perp ab$ above $T_c${, called precursor diamagnetism,} may be parameterized as: 
\begin{eqnarray}
\Delta\chi(T)_H\equiv\frac{M(T,H)}{H}-\frac{M_{B}(T,H)}{H},
\label{eq4}
\end{eqnarray}
where $M(T,H)$ and $M_{B}(T,H)$ are the as-measured and, respectively, background magnetizations. As an example, the $\Delta\chi(\varepsilon)_H$ data obtained from the results of Fig.~\ref{f3}(b) are shown in Fig.~\ref{f5}(b). The corresponding $T_c$ was estimated from the low-field ZFC magnetic susceptibility versus temperature curves, as illustrated in Fig. 3(a), whereas the background susceptibility [the solid line in Fig. 3(b)] was estimated by a linear fit between $160-275$~K (approximately $1.8-3$~$T_c$). We have checked that the resulting $-\Delta\chi(\varepsilon)_H/T$ curve is stable to changes of the background fitting region, and to changes of  the critical temperature between $T_c$ and $T_c\pm\Delta T_c$. These different uncertainties are represented by the shaded area in Fig.~5(b) which, as it was the case for the paraconductivity, around $T_{\rm onset}$ (\textit{i.e.}, $\varepsilon\approx0.5$) also covers the uncertainties estimated for $T_{\rm onset}$ in the precedent subsection.

\begin{figure}[t]
\begin{center}
\includegraphics[scale=.7]{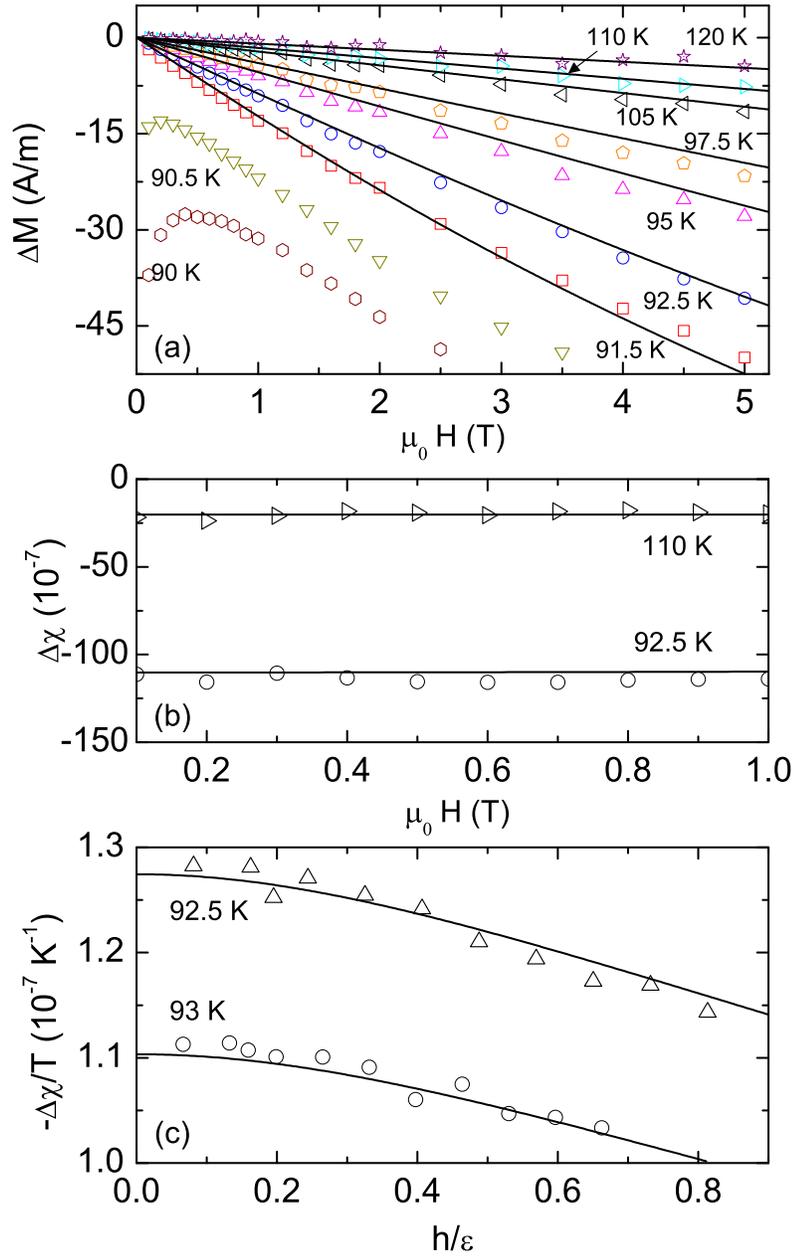}
\caption{(a) Magnetic field dependence of the fluctuation magnetization for several temperatures around $T_c$. The lines were calculated from Eq.~\ref{eq8} by using the same parameters as in the previous analyses (see the main text). (b) Detail to better appreciate the linear behavior of $\Delta\chi(H)$ below 1 T. (c) Examples of the $-\Delta\chi/T$ dependence on $h/\varepsilon$. The lines are the GGL prediction under a total-energy cutoff for the Prange regime [Eq.~\ref{eq8}] with the same $\xi_c(0)$, $\xi_{ab}(0)$ and $\varepsilon^c$ values as in the Schmidt limit. }
\label{f7}
\end{center}
\end{figure}

We have checked at a quantitative level that, as already suggested by the results summarized in Fig.~\ref{f4}, the magnetization above $T_{\rm onset}$ is temperature-independent at the scales of the rounding effects around $T_c$. So, the results presented in Fig.~\ref{f7}(a) for $\Delta M(H)_T$ were obtained by using as background the magnetization isotherm measured at 135 K, which is relatively near $T_{\rm onset}$. A relevant aspect of these isotherms is their linear dependence on $H$ in the so-called low field (or Schmidt, see next section) regime, defined by the condition $H/H_{c2}(0)\ll\varepsilon$. This is further illustrated by the precursor diamagnetism data summarized in Figs. 7(b) and 7(c). One may conclude then that the $\Delta \chi(\varepsilon,H)$ curve plotted in Fig. 5(b), measured under a field amplitude of 1~T, represents well the  $\varepsilon$-dependence of the precursor diamagnetism in the zero-field limit. This will allow the comparison, presented in the next subsection, with the paraconductivity, this last measured in absence of a magnetic field.  

\subsection{The ratio between the precursor diamagnetism and the paraconductivity}

The $\varepsilon$-dependence of the ratio between the precursor diamagnetism in the zero field limit and the paraconductivity may be directly obtained from the results presented in Figs.~5(a) and 5(b). The resulting data points are presented in Fig.~5(c), and they cover the $\varepsilon$-region between 0.02 and 0.4. Above this reduced temperature the results are no longer reliable, due to the combination of the uncertainties of each observable. 

The ratio between $\Delta \chi(\varepsilon)_H$ and $\Delta \sigma_{ab}(\varepsilon)$ has been earlier determined in Ref.~\cite{ref11} (only up to $\varepsilon\approx0.1$) and Ref.~\cite{ref20} (in this case up to $\varepsilon\approx0.4$). Our present data agree at a quantitative level with these earlier results up to $\varepsilon\approx0.1$. However, the increase of the $-\Delta\chi/T\Delta \sigma_{ab}$ ratio for $\varepsilon>0.1$ proposed in Ref.~\cite{ref20} is not confirmed here, where the $\varepsilon$-independent behavior is observed up to the highest accessible reduced temperature. These differences may be attributed to the procedure used to determine the background contribution, which in the present work is more accurate than in Ref.~\cite{ref20}, mainly in the high-$\varepsilon$ region. 

{It is useful to comment already here that on the grounds of the GGL scenario that is going to be detailed in the next Section,} our present results on the $\varepsilon$-dependence of the $-\Delta\chi/T\Delta\sigma_{ab}$ ratio already provide direct, model-independent, information on the superconducting fluctuation effects on $\Delta \sigma_{ab}(\varepsilon)$ and $\Delta \chi(\varepsilon)$: In spite of their very different nature, both observables have, at a quantitative level, the same $\varepsilon$-behavior. This finding excludes possible contributions associated to their specific nature, like indirect fluctuation effects in the case of $\Delta \sigma_{ab}(\varepsilon)$, or the presence of appreciable non-local electrodynamic effects in the case of $\Delta \chi(\varepsilon)$ (see, {\it e.g.}, Ref.~\cite{ref2}). The other central information in Fig.~5(c) is the value itself of the $-\Delta\chi/T\Delta\sigma_{ab}$ quotient, found to be $\simeq3\times10^{-13}~{\rm \Omega mK}^{-1}$. We will see in the next Section that, when combined, this double information provides a crucial experimental check for any theoretical proposal for both the paraconductivity and the precursor diamagnetism. 

%
%

\section{Quantitative analyses of the measured roundings in terms of the GL approach with superconducting fluctuations}

The experimental data summarized in the precedent Section will be first used to probe the $T_{\rm onset}$ associated with the limits imposed by the Heisenberg uncertainty principle to the shrinkage of the superconducting coherence length when the temperature increases above $T_c$. Then, most of this Section will be devoted to probe the mean-field Ginzburg-Landau scenario with superconducting fluctuations around the transition. This comparison will include the extended (with a total-energy cutoff, that takes into account the quantum localization) Gaussian-Ginzburg-Landau (GGL) approach for multilayered superconductors above $T_c$ and, consistently, the GL scaling in the lowest-Landau-level (GL-LLL) approximation for 3D superconductors under finite applied magnetic fields in the critical region around $T_c(H)$.\footnote{The relevance of other contributions to the in-plane paraconductivity and magnetoconductivity in Y-123 (Maki-Thomson, regular Maki-Thomson, and DOS, see Ref.~\cite{DorinDOS}) were studied in detail in Ref.~\cite{RamalloDOS}. It was concluded that there is no consistent set of values for the parameters involved in these contributions that produce a simultaneous agreement with the experiments. Therefore, these contributions do not seem to give a good description of the effects of fluctuations in planar transport, regardless of whether they are useful or not to describe transport in the direction perpendicular to the planes, which is the purpose of the DOS effects, as initially proposed in Ref.~\cite{Ioffe93}. For detailed measurements of the fluctuation conductivity normal to the \textit{ab} layers see Ref.~\cite{Axnas}. Other aspects are commented in subsection 5.1.}
Detailed descriptions of these theoretical results may be seen in the original works.\cite{ref3a,ref3b,ref4,ref5,ref6,ref7a,ref7b,ref16,ref21,ref22,ref23,ref70,ref71} An extension of the GGL calculations to the finite magnetic field regime will be summarized in Appendix A. Here we will just comment briefly on some aspects of the GL scenario that are still controversial, as the influence of the multilayering or the adequacy of the total-energy cutoff.

\subsection{Onset temperature of the superconducting fluctuations}

As earlier discussed in the pioneering work of Gollub and coworkers\cite{ref72} (see also Refs.~\cite{ref2} and \cite{ref6}), the mean-field GGL approaches are formally valid only in the $\varepsilon$-region $\varepsilon_{LG} \leq \varepsilon \ll 1$, where $\varepsilon_{LG}$ is the so-called Levanyuk-Ginzburg reduced temperature. Since then, different attempts have been proposed to extend to high-$\varepsilon$ these mean-field descriptions, including the introduction of different versions of the conventional momentum cutoff,\cite{ref6,ref7a,ref7b,ref8,ref9,ref10a,ref10b,ref10c,ref10d,ref10e,ref10f,ref10g,ref10h,ref10i,ref10j,ref10k,ref10l,ref29,ref30a,ref30b,ref30c,ref30d,ref30e,ref30f,ref30m,ref30o,ref30p,ref30l,ref30g,ref30n,ref30q,ref30r,ref30i,ref30j,ref30k,ref58,ref59,ref60,ref61,ref62,ref63,ref64a,ref64c,ref64b,ref73} or an \textit{ad hoc} penalization (not a cutoff) of the short-wavelength fluctuation modes that already takes into account the quantum localization.\cite{ref74a,ref74b} However, none of these earlier proposals lead to the vanishing of all fluctuation modes above a well-defined temperature, $T_{\rm onset}$. Such a $T_{\rm onset}$ was proposed in Ref.~\cite{ref21} by just taking into account the limits imposed by the Heisenberg uncertainty principle to the shrinkage, when the temperature increases above $T_c$, of the superconducting coherence length, $\xi(T)$, that cannot be smaller than $\xi_{T=0{\rm K}}$, the actual coherence length at zero temperature first introduced by Pippard,\cite{ref2,ref75} \textit{i.e.}, 
\begin{eqnarray}
\xi(T) \stackrel{>}{_\sim} \xi_{T=0{\rm K}}.
\label{eq6}
\end{eqnarray}
This condition directly leads to a well-defined $T_{\rm onset}$, given by $\xi(T_{\rm onset})=\xi_{T=0{\rm K}}$, above which all fluctuation modes vanish, independently of their wavelength. The corresponding cutoff reduced temperature is then just $\varepsilon^c\equiv$ ln$(T_{\rm onset}/T_c)$. Note also that the above condition is indeed compatible with the existence of superconductivity in samples with sizes smaller than the value of $\xi_{T=0{\rm K}}$ in the bulk. This is because, as earlier commented elsewhere (second paper in Ref.~\cite{ref21}), in such small superconductors the Pippard coherence length looses its conventional meaning (from a crude point of view, in these small superconductors the coherence length amplitude is also reduced with respect to the bulk value).

As also stressed in Ref.~\cite{ref21}, the above condition is general, and must apply to any theoretical description of the superconducting transition.  Only the value of $\varepsilon^c$ will depend, through the temperature dependence of $\xi(T)$ and the relationship between $\xi(T)$ and $\xi_{T=0{\rm K}}$, of each particular approach. A relevant example will correspond to the combination of the mean-field temperature dependence of the coherence length, $\xi(T)=\xi(0)\varepsilon^{-1/2}$ [and then $\varepsilon^c=(\xi(0)/\xi_{T=0{\rm K}})^2$], with the relationship between $\xi(T)$ and  $\xi_{T=0{\rm K}}$ proposed by the mean-field BCS theory, which in the clean limit is $\xi(0)$= 0.74$\xi_{T=0{\rm K}}$.\cite{ref75} This leads then to $\varepsilon^c\approx$0.55, \textit{i.e.}, $T_{\rm onset}\approx$1.7$T_c$,  in excellent agreement with our present experimental results for the onset of the resistivity and the magnetization roundings in the zero-field limit presented in Figs.~\ref{f1}(c) and, respectively, \ref{f3}(c), and summarized in Table I. 
 
\begin{figure}[t]
\begin{center}
\includegraphics[scale=.7]{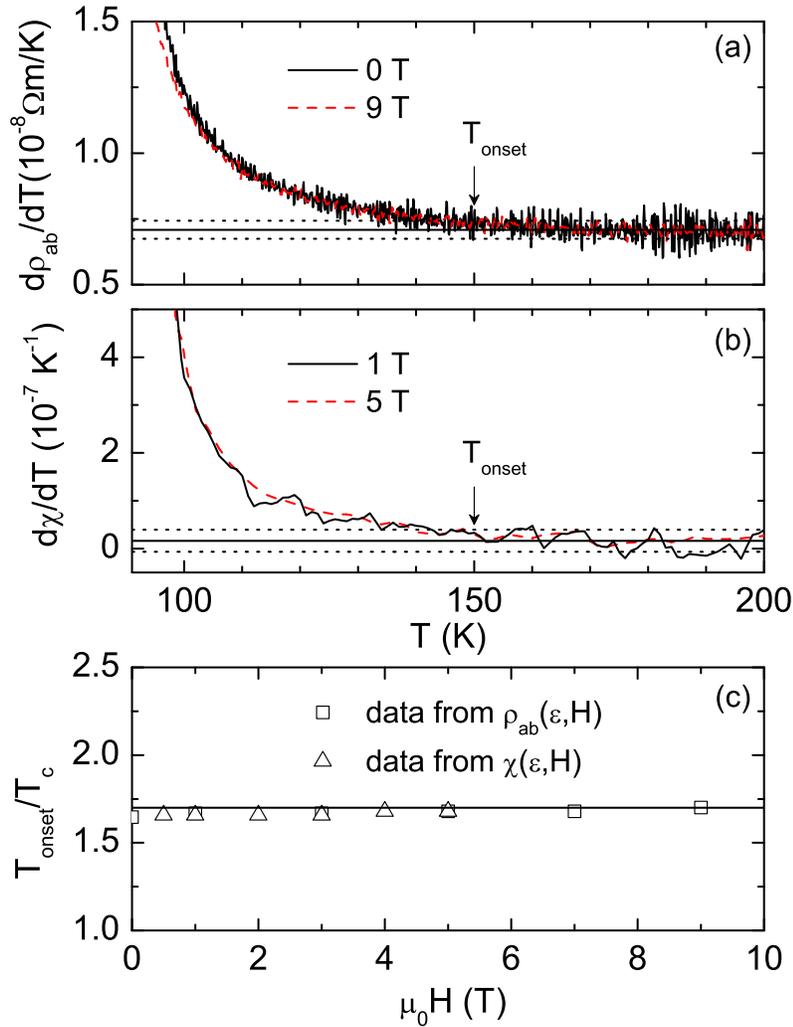}
\caption{(a)-(b) Temperature dependence of $d\rho_{ab}/dT$ and $d\chi/dT$ from measurements under two different magnetic fields. These data illustrate that $T_{\rm onset}$ is field independent, up to 9~T and, respectively, 5~T. The dotted lines delimit the noise level (see the main text for details). (c) $T_{\rm onset}/T_c$ for the two observables as a function of the applied magnetic field amplitude. The solid line corresponds to $T_{\rm onset}/T_c=1.7$. }
\label{f8}
\end{center}
\end{figure}

The total-energy cutoff reduced-temperature, $\varepsilon^c$, may also be affected by the application of a magnetic field large enough as to reduce the superconducting coherence length amplitude, {\it i.e.}, of the order of $H_{c2}(0)$.\cite{ref21} This effect was observed experimentally in low-$T_c$ superconductors,\cite{ref76} and in a low-$T_c$ cuprate.\cite{lowTcHTSC} However, in our experiments the maximum field amplitude was 9~T, much lower than the $\mu_0H_{c2}(0)$ value for OPT Y-123, and thus one might expect that $\varepsilon^c$ will remain field independent. The results summarized in Fig.~\ref{f8} fully confirm this prediction. Other aspects of $\varepsilon^c$ will be commented in the next Section.
 
%
%

\subsection{Comparison of the fluctuation-induced conductivity, magnetoconductivity and diamagnetism in the mean-field region above $T_c$ with the GGL model equations\label{subsec-comparisonGGL}}

As stressed above, the experimental finding that the ratio between the precursor diamagnetism and the paraconductivity does not depend on temperature, a model-independent result, strongly suggests the absence of indirect fluctuation effects on the electrical conductivity and of non-local electrodynamic effects on the precursor diamagnetism. These effects may be then neglected, providing a first simplification when analyzing the experimental data. Another relevant simplification concerns the effective distance, $s$, between adjacent superconducting $ab$-layers. As explained in detail in Refs.~\cite{ref5,ref6,ref35}, in the GGL approach above $T_c$ in bilayered superconductors with two alternate interlayer distances $s_1$ and $s_2$ (such as Y-123) the effective $s$ becomes equal to the bilayer periodicity $s_1+s_2$ only if the corresponding Josephson interlayer couplings, $\gamma_1$ and $\gamma_2$, differ among them by orders of magnitude, \textit{i.e.}, $\gamma_1/\gamma_2\to\infty$. Instead, $\gamma_1/\gamma_2\stackrel{<}{_\sim}30$ in OPT Y-123, and in this case $s=(s_1+s_2)/2$ is an excellent approximation.\cite{ref22,ref77}. Then, in what follows we will use $s=(s_1+s_2)/2$ as effective interlayer distance. The detailed analyses presented below will fully confirm the adequacy of these two starting simplifications.

To probe the GGL approach we will then use the expressions for the direct fluctuation effects on the in-plane paraconductivity and the precursor diamagnetism in layered superconductors in a finite field $H\perp ab$, and under a total-energy cutoff. The calculations of these expressions are summarized in Appendix~A. The in-plane paraconductivity is given by:
\begin{equation}
\hspace{-1cm}\Delta\sigma_{ab}(\varepsilon,h)=\frac{\tau_{\rm rel} e^2}{64\hbar \pi h}\int_{-\pi/s}^{\pi/s} 
\left[\psi^{(1)}\left(\frac{h+\varepsilon+\omega_{k_z}}{2h}\right)-\psi^{(1)}\left(\frac{h+\varepsilon^c+\omega_{k_z}}{2h} \right)
 \right] \,\mathrm{d}k_z,
\label{eq7}
\end{equation}
whereas  the precursor diamagnetism is:
\begin{eqnarray}
\frac{\Delta \chi(\varepsilon,h)}{T}
&=&\frac{-k_B}{2 \pi \phi_0 h \textsl{H}_{c2}(0)}
\int_{-\pi/s}^{\pi/s} 
\left[-\frac{\varepsilon^c+w_{k_z}}{2h} \psi\left(\frac{h+\varepsilon^c+w_{k_z}}{2h}\right)\right.\nonumber\\ 
&&\left.-\ln \Gamma\left(\frac{h+\varepsilon+w_{k_z}}{2h}\right)
+\ln\Gamma \left(\frac{h+\varepsilon^c+w_{k_z}}{2 h}\right)\right.\nonumber\\ 
&&\left.+\frac{\varepsilon+w_{k_z}}{2h}\psi\left(\frac{h+\varepsilon+w_{k_z}}{2h}\right)+\frac{\varepsilon^c-\varepsilon}{2h}\right] \,\mathrm{d}k_z .  
\label{eq8}
\end{eqnarray}
Here $\tau_{\rm rel}\equiv\tau_0(0)/\tau_0^{BCS}$ is the GL amplitude of the relaxation time of the fluctuating Cooper pairs for a wave vector ${\bf k} = 0$, expressed in adimensional units relative to the BCS amplitude $\tau_0^{BCS}=\pi\hbar/(8k_BT_{c})$,\cite{ref78} $e$ is the electron charge, $\hbar$ is the reduced Planck constant, $k_B$ is the Boltzmann constant, $\phi_0$ is the magnetic flux quantum,  $h \equiv H/H_{c2}(0)$ is the reduced magnetic field, $w_{k_z}=B_{LD}(1-\cos(k_zs))/2$ is the out-of-plane spectrum of the fluctuations, $B_{LD}=(2\xi_c(0)/s)^2$ is the so-called Lawrence-Doniach (LD) parameter, $\Gamma$, $\psi$ and $\psi^{(1)}$ are the gamma, digamma and trigamma functions, and $\xi_c(0)$ and $\xi_{ab}(0)$ are the amplitudes of the transverse and in-plane GL coherence lengths. As indicated above, for the superconducting layers' periodicity length we will use $s = (s_1+s_2)/2=0.59$~nm. In principle the total-energy cutoff constant, $\varepsilon^c$, could be  directly determined from the $T_{\rm onset}$ obtained in the previous subsection. However, as a further check of consistency,   here we will leave $\varepsilon^c$ free and  we will compare the result with $\ln(T_{\rm onset}/T_c)$. 

In the low magnetic field limit (\textit{i.e.}, for $h\ll\varepsilon,\varepsilon^c$) Eqs.~\ref{eq7} and (\ref{eq8}) reduce to
\begin{equation}
\Delta\sigma_{ab}(\varepsilon)=\frac{\tau_{\rm rel} e^2}{16\hbar s}\left[\frac{1}{\varepsilon}\left(1+\frac{B_{LD}}{\varepsilon}\right)^{-1/2}
-\frac{1}{\varepsilon^c}\left(1+\frac{B_{LD}}{\varepsilon^c}\right)^{-1/2}\right],
\label{eq9}
\end{equation}
and, respectively,
\begin{equation}
\frac{\Delta \chi(\varepsilon)}{T}=-\frac{\pi k_B\mu_0\xi_{ab}^2(0)}{3 \phi_0^2 s}\left[\frac{1}{\varepsilon}\left(1+\frac{B_{LD}}{\varepsilon}\right)^{-1/2}
-\frac{1}{\varepsilon^c}\left(1+\frac{B_{LD}}{\varepsilon^c}\right)^{-1/2}\right],
\label{eq10}
\end{equation}
where $\mu_0$ is the vacuum permeability. In turn, for $\varepsilon\ll\varepsilon^c$ (that  corresponds then to a region outside the short-wavelength fluctuation regime), Eqs.~\ref{eq9} and (\ref{eq10}) reduce to
\begin{equation}
\Delta\sigma_{ab}(\varepsilon)=\frac{\tau_{\rm rel} e^2}{16\hbar s}\frac{1}{\varepsilon}\left(1+\frac{B_{LD}}{\varepsilon}\right)^{-1/2},
\label{eq11}
\end{equation}
that is the well-known Aslamazov-Larkin (AL) expression for the paraconductivity in the LD scenario,\cite{ref79} and to
\begin{eqnarray}
\frac{\Delta \chi(\varepsilon)}{T}=-\frac{\pi k_B\mu_0\xi_{ab}^2(0)}{3 \phi_0^2 s}\frac{1}{\varepsilon}\left(1+\frac{B_{LD}}{\varepsilon}\right)^{-1/2},
\label{eq12}
\end{eqnarray}
that is the GGL prediction for the fluctuation diamagnetism in the Schmidt limit for layered superconductors, first proposed by Lawrence and Doniach\cite{ref79} (and, independently, by Tsuzuki\cite{ref80} and by Yamaji\cite{ref81}).

Another relevant result of the GGL scenario, which may be directly obtained by just combining Eqs.~\ref{eq9} and (\ref{eq10}), is that the relationship between the precursor diamagnetism and the paraconductivity in the low field (Schmidt) limit is temperature-independent,
\begin{eqnarray}
\frac{-\Delta\chi(\varepsilon)}{T\Delta\sigma_{ab}(\varepsilon)}= \frac{16\mu_0k_B}{3\pi\hbar} \frac{\xi_{ab}^2(0)}{\tau_{\rm rel}}=2.79\times10^5\frac{\xi_{ab}^2(0)}{\tau_{\rm rel}}\; {\rm (SI\,units)},
\label{eq13}
\end{eqnarray}
a result that extends to the high-$\varepsilon$ region our earlier proposal.\cite{ref11,ref20} \footnote{The relationship between $\Delta \chi(\varepsilon)_H$ and $\Delta \sigma_{ab}(\varepsilon)$, first proposed in  Ref.~\cite{ref11}, was since then applied by different authors to analyze the superconducting fluctuations around $T_c$ in different high-$T_c$ superconductors (see, {\it e.g.},  Ref.~\cite{ref6}). Examples of recent works using such a relationship are the second work in Ref.~\cite{ref30l}, and Refs.~\cite{ref30i,ref30k}.} Equation~(\ref{eq13}) agrees with the experimental results summarized in Section 3.5 and in Fig.~\ref{f5}(c). This agreement further suggests the absence of indirect fluctuation effects on the paraconductivity and of non-local electrodynamic effects on the precursor diamagnetism.\cite{ref2} The solid line in Fig.~\ref{f5}(c) corresponds to the best fit of Eq.~\ref{eq13} to the data, with $\xi_{ab}^2(0)/\tau_{\rm rel}$ as the only free parameter. This leads to $\xi_{ab}^2(0)/\tau_{\rm rel}\simeq$ 1.1 nm$^2$ a value that provides a relevant constraint to the remaining free parameters of the GGL scenario.

In order to disentangle the values of $\tau_{\rm rel}$ and $\xi_c(0)$, and also to determine the remaining free parameters entering the GGL expressions, {\it i.e.}, $\xi_c(0)$ and $\varepsilon^c$, we need to simultaneous and consistently fit Eqs.~(\ref{eq7}) to (\ref{eq10}) to our full set of $\mbox{$\Delta\sigma_{ab}(\varepsilon)$}$, $\mbox{$\Delta\sigma_{ab}(\varepsilon,H)$}$, $\Delta\chi(\varepsilon)$ and  $\Delta M(\varepsilon,h)$ data. The combined fit to all our data above $\varepsilon=0.02$ is presented in Figs.~5 to 7. The values for the fitting parameters are summarized in Table~II. 

%
%
\begin{table}[h]
\begin{center}
\begin{tabular}{cccc}
\hline
$\xi_c(0)$  &  $\xi_{ab}(0)$ & $\tau_{\rm rel}$ & $\varepsilon^c$ \\
(nm)  &  (nm) &  &  \\
\hline 
$0.11\pm0.02$ &  $1.1\pm0.2$ & $1.0\pm0.2$  & $0.55\pm0.15$ \\
\hline
\end{tabular}
\end{center}
\caption{Values of the parameters arising in the mean-field Ginzburg-Landau scenario. With these values, the GL approaches explain at a quantitative level, simultaneous and consistently, the precursor diamagnetism in the low magnetic field limit (Schmidt regime) and under moderate fields (Prange regime), the paraconductivity and the fluctuation-induced magnetoconductivity.}
\end{table}

It may be illustrative to discuss some of those observables separately in specific $(\varepsilon,h)$ ranges of particular interest. For instance, Fig.~5(a) shows that for the zero-field paraconductivity $\mbox{$\Delta\sigma_{ab}(\varepsilon)$}$ it is not necessary to take cutoff effects into account for $\epsilon\stackrel{<}{_\sim}0.1$, where Eq.~\ref{eq11} (dashed line in that figure) accounts for the data. This same equation fails, however, to account for the rapid fall of fluctuations at larger $\varepsilon$-values, for which the use of Eq.~\ref{eq9} is needed (continuous line in the same figure). A similar situation happens for the low-field fluctuation magnetic susceptibility, displayed in Fig.~5(b): While  for $\epsilon\stackrel{<}{_\sim}0.1$ Eq.~\ref{eq12} without a cutoff accounts for the data, at larger $\varepsilon$ values Eq.~\ref{eq10} is needed.

Concerning the effects of a finite magnetic field on the fluctuation-induced observables, Figs.~6 and 7 illustrate the $H$-dependence of both $\mbox{$\Delta\sigma_{ab}(\varepsilon,H)$}$ and $\Delta M(\varepsilon,h)$. It is evident in these figures that Eqs.~\ref{eq7} and (\ref{eq8}) account very well for these data, using the parameter values summarized in Table~II.

%
%
\begin{figure}[t]
\begin{center}
\includegraphics[scale=.7]{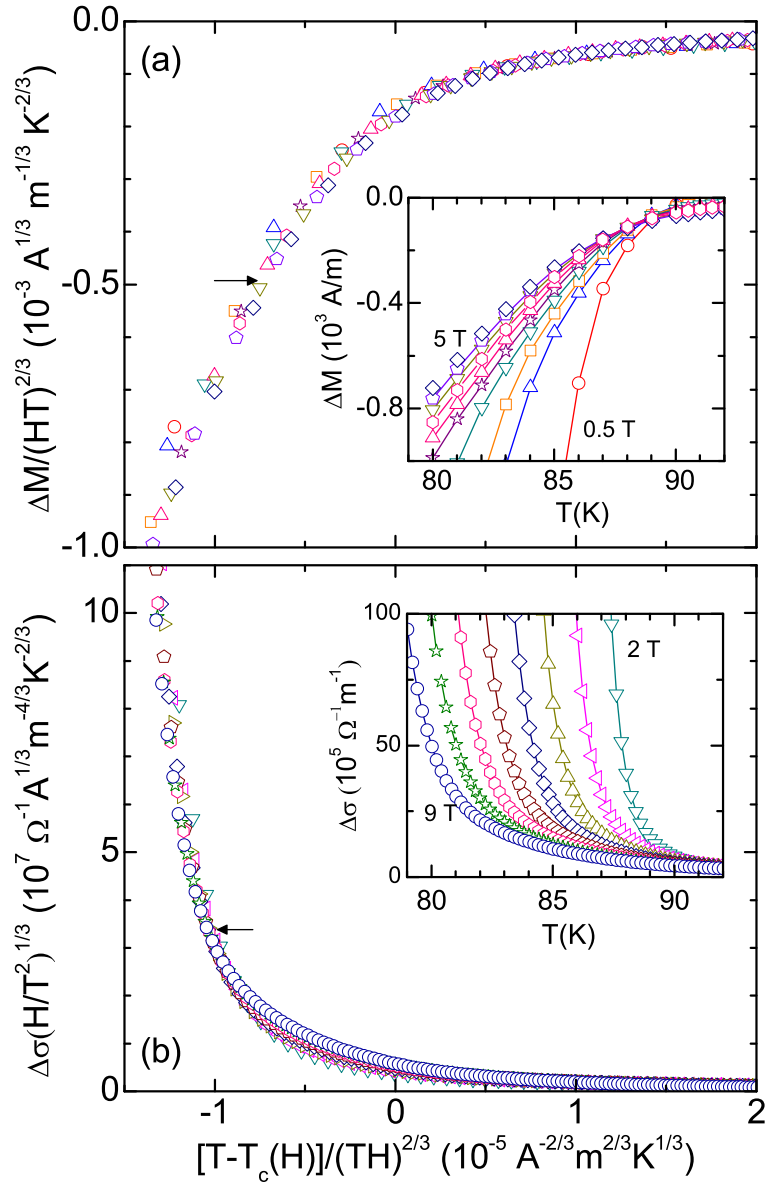}
\caption{(a)-(b) 3D-LLL scaling of the magnetization (for magnetic fields between 0.5~T and 5~T in steps of 0.5~T) and of the fluctuation conductivity (for magnetic fields between 2~T and 9~T in steps of 1~T) for temperatures around $T_c(H)$. The scaling variables were evaluated by using the superconducting parameters resulting from the analysis in the Gaussian region. The arrows indicate the limit of applicability of the scalings. Insets: Temperature dependence of the raw fluctuation magnetization and conductivity.}
\label{f9}
\end{center}
\end{figure}

\subsection{GL scaling of the fluctuation diamagnetism and conductivity in the critical region around $T_c(H)$}

Although the analysis of the rounding effects in OPT Y-123 in terms of the GL scenario presented in this paper focuses on the behavior in the Gaussian mean-field-like region above $T_c$, and they were done on the grounds of the {extended} GGL approximation, for completeness in this subsection we will analyze the critical region around the $T_c(H)$ line, where the Gaussian approximation is no longer valid. For that we will use the GL approach for 3D superconductors proposed by Ullah and Dorsey,\cite{ref4} based on the lowest-Landau-level (LLL) approximation. The adequacy of this GL-LLL approach to explain the  fluctuation diamagnetism and conductivity observed around $T_c$ in OPT Y-123 was also earlier demonstrated\cite{ref13a,ref13b,ref13c,ref13d,ref13e} (see also Ref.~\cite{ref83}). In the framework of this approach, $\Delta M(T,H)$ and  $\Delta\sigma_{ab}(T,H)$ follow a scaling behavior,\cite{ref4}
\begin{eqnarray}
\Delta M(T,H)_{3D}=(TH)^{2/3}F_M\left(\frac{T-T_c(H)}{(TH)^{2/3}}\right)
\label{eq14}
\end{eqnarray}
and
\begin{eqnarray}
\Delta \sigma_{ab}(T,H)_{3D}=\left(\frac{T^2}{H}\right)^{1/3}F_\sigma\left(\frac{T-T_c(H)}{(TH)^{2/3}}\right),
\label{eq15}
\end{eqnarray}
where $F_M$ and $F_\sigma$ are scaling functions.\cite{ref84} In applying Eqs.~\ref{eq14} and (\ref{eq15}) to the experimental data, $T_c(H)$ was determined from the GL relation 
\begin{eqnarray}
T_c(H)=T_c\left[1-\frac{H}{\phi_0/2\pi\mu_0\xi_{ab}^2(0)}\right],
\label{eq17}
\end{eqnarray}
by using the $T_c$ and $\xi_{ab}(0)$ values resulting from the analysis in the Gaussian region. The resulting scalings are shown in Figs.~9(a) and 9(b), where the corresponding unscaled $\Delta M(T,H)$ and $\Delta \sigma_{ab}(T,H)$ data are also presented in the insets. As it may be seen, the scaling is excellent for both observables down to $(T-T_c(H))/(TH)^{2/3}\sim -10^{-5}$ ${\rm A}^{-2/3}{\rm m}^{2/3}{\rm K}^{1/3}$ (indicated by an arrow),  in agreement with previous works.\cite{ref13a,ref13b,ref13c,ref13d,ref13e,ref83} The present results nicely extend the applicability of the GL-LLL approaches to fields as large as 9~T, and represent a stringent consistency check of the analysis in the Gaussian region. It is also worth noticing the existence of self-consistent calculations of $\Delta M$ and $\Delta\sigma$ in layered superconductors based on the GL-LD model,\cite{Puica2003,ref64b,Wang2016} that agree with the experimental data near $T_c(H)$.\cite{ref64b,Wang2016}

\subsection{Phase diagram for the superconducting fluctuations {scenario}}

A schematic $H-T$ phase diagram {of the GL scenario showing the different regions for the superconducting fluctuations is presented in Fig.~10. The regions covered by our measurements of the rounding} effects on the electrical conductivity and magnetization (shaded area) is presented in Fig.~10(a). In Figs.~10(b) and 10(c) the color scale represents the fluctuation-induced conductivity. In Fig.~10(b), the diamonds  show the experimental onset temperature for each field and the solid line near them is the estimate $\xi_{ab}(T_{\rm onset},H)=\xi_{T=0{\rm K}}$ (that for the low fields in this figure corresponds to $T_{\rm onset}\approx$ 1.7$T_c$). To better appreciate  the critical region, Fig.~10(c) shows a detail centered on $T_c(H)$. 
In this figure, squares and circles represent the lower limit of applicability of the GL-LLL scalings of $\Delta M(T,H)$ and, respectively, $\Delta\sigma_{ab}(T,H)$. The critical region boundary (dashed line) was evaluated by fitting to these data the behavior predicted by the so-called $H$-dependent Ginzburg criterion for 3D superconductors,\cite{Ikeda}
\begin{equation}
\frac{\left| T-T_c(H) \right|}{T_c}\propto\left(\frac{H}{H_{c2}(0)}\right)^{2/3}.
\label{eq16}
\end{equation}
Within the Gaussian region the dotted line represents the condition $h\approx0.2\varepsilon$, roughly separating the Prange regime from the low-$H$ Schmidt regime. For completeness, we also indicate the experimental irreversibility line (triangles) obtained from the temperatures at which $\rho_{ab}(T)_H=0$ [see Fig.~2(b)].  

%
%
\begin{figure}[t]
\begin{center}
\includegraphics[scale=.55]{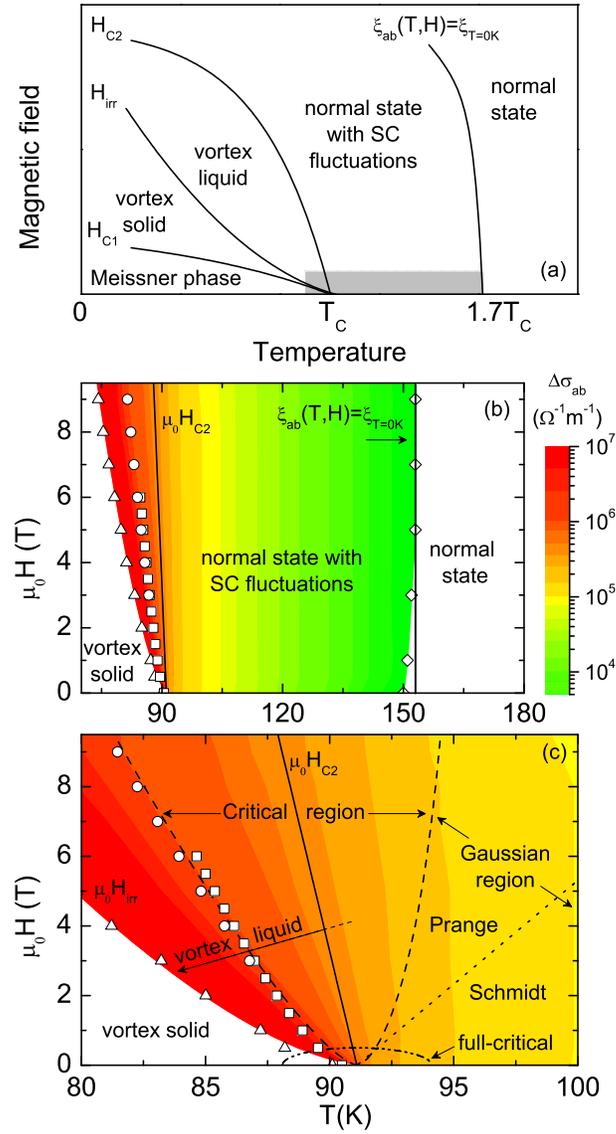}
\caption{(a) Schematic $H-T$ phase diagram {of the GL scenario with superconducting fluctuations showing the regions covered by the measurements summarized in this paper (shaded area).} The color scale in (b) represents the fluctuation-induced conductivity, and the diamonds show the onset temperature for each field (the solid line over them is $T_{\rm onset}\approx$ 1.7$T_c$). A detail around $T_c$ is presented in (c). Circles and squares are the limit of applicability of the GL-LLL scalings of $\Delta\sigma_{ab}$ and, respectively, $\Delta M$. The dashed line is a fit to the $H$-dependent Ginzburg criterion, Eq.~\ref{eq16}. The solid line is the upper critical field given by  Eq.~\ref{eq17}. The dotted line in the Gaussian region corresponds to $h\approx0.2\varepsilon$, and roughly separates the Prange regime and the low-$H$ region (Schmidt regime). The irreversibility line (triangles) was obtained from the temperatures at which $\rho_{ab}(T)_H=0$. Closer to $T_c$ and in the zero field limit is the so-called full-critical region, indicated by the dashed-dotted line. }
\label{f10}
\end{center}
\end{figure}

It is worth noting that direct application of Eq.~(\ref{eq16}) leads to a null width of the critical region in the limit $H\to0$. This  arises from the failure at  low fields of the LLL approximation used by Eq.~(\ref{eq16}). At $H=0$, the width of the  full-critical region (where the mean-field-like approaches are no longer applicable and the superconducting fluctuations are described instead in terms of the XY model) may be approximated by the well-known Ginzburg criterion originally calculated for 3D superconductors by Levanyuk\cite{Levanyuk} and Ginzburg\cite{Ginzburg} (for an extension to layered superconductors, see Ref.~\cite{ref85}). The precise field-dependence of the critical region at low fields, and how it merges with the low-field full-critical region, is still unknown. Other still open issues concerning the full-critical region at zero field include the behavior of the fluctuation conductivity and magnetization: As already mentioned, the uncertainties in the determination of $T_c$ (and also the  large effect of even a small $T_c$ distribution due to inhomogeneities, unavoidable in real samples) difficult determining the bare critical exponents in the very close vicinity of $T_c(H=0)$. In moderately anisotropic superconductors such as OPT Y-123, the 3D~XY model is expected to provide a first  approximation to the fluctuations in the full-critical region. But it could also be expected that the layered structure may induce  extra fluctuation contributions arising from vortex-antivortex effects akin to those first proposed by Berezinskii\cite{Berezinskii} and by Kosterlitz and Thouless\cite{KT} for 2D superfluids (for a theoretical discussion of the effect in layered superconductors see, {\it e.g.}, Refs.~\cite{Shenoi, Buzdin, ref42}, and for a measurement in the much more anisotropic La$_{2-x}$Sr$_x$CuO$_4$ see Ref.~\cite{Coton}). For OPT Y-123, partial evidences of 3D~XY exponents in the zero-field fluctuation conductivity for $\varepsilon<0.02$ have been in fact early reported in, {\it e.g.},  Ref.~\cite{ref9}, while some indications of a behaviour of the Berezinskii-Kosterlitz-Thouless type were suggested by measurements in, {\it e.g.},  Ref.~\cite{Ying}. Further work will be clearly needed to fully understand these aspects.

\section{On the application to the rounding effects measured in the OPT Y-123 of the scenarios associated with the presence of charge density waves and with the pseudogap.}

As already noted in the Introduction, different unconventional (non-Ginzburg-Landau like) scenarios have been proposed for the  rounding effects measured above $T_c$ in cuprate superconductors, the most popular being those based on the possible existence of a region of phase disordering up to the temperature where a pseudogap opens.\cite{ref24,ref26,ref27,ref28,ref29,ref30a,ref30b,ref30c,ref30d,ref30e,ref30f,ref30m,ref30o,ref30p,ref30l,ref30g,ref30n,ref30q,ref30r,ref30i,ref30j,ref30k,ref31,ref32,ref33,ref34,ref35,ref36,ref37a,ref37b,ref37c,ref38,ref39,ref40,ref41,ref42,ref43,ref44,ref45,ref46,ref47,ref48,ref49,ref50,ref51,ref52}  Unfortunately, to the best of our knowledge, up to now the theoretical approaches to account for these rounding effects in terms of non-GL scenarios are still not well settled. Nevertheless, for completeness in this Section we will briefly present some comments on the role that could play on these effects the possible presence of charge density waves (CDW) and also the pseudogap opening. In this last case, we have already commented, in particular in Refs.~\cite{ref32,ref34,ref35,ref36}, various aspects of their confrontation with the Gaussian-Ginzburg-Landau (GGL) scenarios for the conventional, thermodynamic-like, superconducting fluctuations. {Let us stress already that our comments here will mainly address the application of these scenarios to the rounding effects around $T_c$ in OPT Y-123 and, therefore, our conclusions will not exclude the relevance of these unconventional effects in overdoped and underdoped HTSC. In fact, the entanglement between superconducting fluctuations and pseudogap effects {could} arise in many of the rounding behaviors measured above $T_c$ in underdoped HTSC, enhancing the interest of our present results when using the resistivity roundings to locate $T^*$ in these compounds.}

\subsection{On the presence of charge density waves}

In what concerns the CDW, as shown, e.g., in Ref.~\cite{Machida81}, their coexistence and mutual interaction with the superconducting order may be described in terms of two simultaneous order parameters, namely, the homogeneous superconducting order parameter associated with the superconducting gap and the periodic order parameter characteristic of the CDW-state. However, for the CDW to have a significant influence on the SC-state, several conditions are required. First of all, if the wave vector of the CDW differs significantly from $2k_F$, their influence on the stability of the Cooper pairs pairing becomes negligible.\cite{Machida81} Once this constrain is fulfilled, the possible effects of the CDW on the superconducting phase can be described by means of a modified Ginzburg-Landau functional that just differs from the conventional one by the presence of a coupling term between both order parameters. A rough estimate of the influence of the CDW on the superconductivity can be done through the first order correction of a perturbative treatment of such coupling term. For $H\perp ab$ this results in a renormalization of the reduced temperature given by\cite{Soto07}
\begin{equation}
\varepsilon\to\varepsilon+K\exp\left[-\frac{Q^2\xi_{ab}^2(0)}{4h}\right].
\end{equation}
Here $K$ is a coupling constant expected to be of the order of magnitude of $T_c/T_{CDW}$, and $Q$ is the wavevector of the CDW. For OPT Y-123 $Q\approx0.3$ (r.l.u.)$\approx5$~nm$^{-1}$,\cite{Blanco14} and $\xi_{ab}(0)\approx1$~nm. Subsequently, the exponential in the above equation is negligible even for the highest reduced magnetic field used in our experiments ($h\approx0.15$). This suggests the absence of appreciable effects of the CDW on the rounding observed in the range of temperatures and magnetic fields accessible in our experiments (see also the comments in the introduction of Section 4).

\subsection{On the presence of a pseudogap}

In what concerns the influence of the possible opening of a pseudogap on the measured rounding effects above $T_c$, the most popular scenarios are those based on the existence of a region of phase disordering up to $T^*$, the temperature where a pseudogap opens.\cite{ref24,ref26,ref27,ref28,ref29,ref30a,ref30b,ref30c,ref30d,ref30e,ref30f,ref30m,ref30o,ref30p,ref30l,ref30g,ref30n,ref30q,ref30r,ref30i,ref30j,ref30k,ref31,ref32,ref33,ref34,ref35,ref36,ref37a,ref37b,ref37c,ref38,ref39,ref40,ref41,ref42,ref43,ref44,ref45,ref46,ref47,ref48,ref49,ref50,ref51,ref52} These proposals are still at present a debated issue and, as noted before, in Refs.~\cite{ref32,ref34,ref35,ref36} we have commented in some detail various aspects of their confrontation with the GGL scenarios. These previous comments include the role that could play in some measurements the presence of chemical inhomogeneities (see also, Ref.~\cite{ref33}). So, to complement these analyses, we will focus here on two particular but interesting points: i) First, the seemingly inaccessibility of the upper critical field line, $H_{c2}(T)$, that is in contrast with the case of conventional low-$T_c$ superconductors. In a Reply (see the second paper in Ref.~\cite{ref27}) to one of our Comments,\cite{ref35} this behavior has been presented as a crucial new support of the phase fluctuation scenario (see also Ref.~\cite{ref47}). ii) Secondly, the implications of our present results indicating that superconducting fluctuations in OPT Y-123 are conventional, GL-type, and that fluctuating superconducting pairs do not exist above $\sim2T_c$ at OPT doping. 

The first point noted above was then the absence in the cuprates, in particular in OPT Y-123, of the characteristic changes in slope in the magnetization versus temperature curves that may be observed in low-$T_c$ superconductors when approaching the superconducting transition. In Ref.~\cite{ref27}, these changes were claimed to be an universal fingerprint of the conventional (Ginzburg-Landau like) superconducting transitions at the $H_{c2}(T)$ line. However, the differences between high-$T_c$ cuprates and low-$T_c$ metallic superconductors may be easily attributed to the important differences between their superconducting parameter values, differences which, as is well known, lead already to a much important amplitudes of the own conventional, thermodynamic like, superconducting fluctuations (fluctuations of both the phase and the amplitude of the superconducting order parameter). A direct consequence is that the critical region around the $H_{c2}(T)$ line (where the GGL approximation is no longer valid) is much wider in high-$T_c$ cuprates. In fact, these differences were earlier accounted for in Ref.~\cite{ref4} as a consequence of the differences between the superconducting parameter values: By using a self-consistent Hartree approximation to treat the quartic term in the Lawrence-Doniach {free energy functional}, these authors showed that, in contrast with the low-$T_c$ superconductors, in the cuprate superconductors the transport and thermodynamic observables must interpolate smoothly through the $H_{c2}(T)$ line (see also Ref.~\cite{ref64b}).

We comment now on the second point indicated above, the implications of our results on the much debated issue of the so-called pseudogap line of the temperature-doping phase diagram of HTS.  First of all, we have to note that various criteria have been suggested for the determination of the pseudogap temperature $T^*$, leading to significantly different results for OPT doping. In particular, some authors (see, e.g., Refs.~\cite{ref30c,Shekhter13,Wu15,Badoux16} and references therein) suggest that the pseudogap temperature would lie well below $T_c$ in OPT Y-123, while other authors (see, e.g., Refs.~\cite{ref1a,ref1b,ref27,ref30g,taillefer,ref47,Xu00}) locate $T^*$ in OPT Y-123 well above $T_c$ (near $1.7T_{c, \rm OPT}$, i.e., close to our $T_{\rm onset}$).

Evidently, the first scenario ($T^*<T_c$ for OPT doping) is straightforwardly compatible with our present results, as the pseudogap physics at that doping level would not be related to, or entangled with, superconducting fluctuations.  This localization in OPT Y-123 of $T^*$ inside the superconducting dome would suggest a pseudogap  not directly related to the superconducting phase, with the corresponding implications for the pairing mechanisms in cuprates.\cite{ref1a,ref1b,ref24,ref25,ref26,ref27,ref28,ref29,ref30a,ref30b,ref30c,ref30d,ref30e,ref30f,ref30m,ref30o,ref30p,ref30l,ref30g,ref30n,ref30q,ref30r,ref30i,ref30j,taillefer} For instance, in Ref.~\cite{Shekhter13}, that finds $T^*<T_c$ for OPT doping by using high-resolution resonant ultrasound spectroscopy, it is proposed that such result would confirm a quantum-critical origin for both the strange metallic behavior observed above the pseudogap line in the normal state and the superconducting pairing mechanism.

Concerning the second scenario ($T^*>T_c$ for OPT doping), our present results provide an interesting thermodynamic constraint for the corresponding explanations of the pseudogap. In particular, our results (like, importantly, the constant ratio $-\Delta\chi/T\Delta\sigma_{ab}$, the accordance of $T_{\rm onset}$ with the value $1.7T_c$, or the excellent agreement of $\Delta\sigma(\varepsilon)$, $\Delta\tilde\sigma(\varepsilon,H)$, $\Delta\chi(\varepsilon)$ and $\Delta M(\varepsilon,H)$ with the quantitative predictions of the GGL and GL-LLL  equations) provide compelling evidence that, in OPT Y-123, the roundings above the transition are already accounted for by fluctuations conventional for both the phase and the amplitude of the superconducting order parameter.  This strongly suggests that any uncoventional fluctuation scenario for underdoped cuprates should evolve as doping approaches the OPT level, in particular recovering a conventional GGL fluctuation behaviour at OPT doping. Fluctuation-pseudogap theories should be, therefore, worked out to introduce this new constraint. In fact, this has been done in part in works such as Refs.\cite{Levin,Geshkenbein,Ussishkin02} that include variations of  strong phase fluctuation theories for the pseudogap that have the capability to recover GGL-type fluctuations in the OPT doping limit.

Let us also briefly dicuss on an additional difficulty affecting the conclusions of some authors that define the pseudogap temperature as the temperature onset for the resistivity rounding above $T_c$ (see, e.g., Ref.~\cite{taillefer} and references therein). 
As is well known, the application of such a criterion is a hazardous task when the in plane resistivity rounding onset manifests, as it is the case of the OPT Y-123, between $T_c$ and 2$T_c$, because of the unavoidable presence of appreciable superconducting fluctuation effects in this temperature region.\cite{ref2,ref3a,ref3b,ref4,ref5,ref6,ref7a,ref7b,ref8,ref9,ref10a,ref10b,ref10c,ref10d,ref10e,ref10f,ref10g,ref10h,ref10i,ref10j,ref10k,ref10l,ref11} 
In fact, under this criterion, $T^*$ in OPT Y-123 will coincide also with the temperature onset of the rounding effects above $T_c$ in other observables, including the magnetization and the Nernst coefficient. An illustrative example for this last observable may be seen in Fig.~2 of Ref.~\cite{taillefer}.  By comparing the interesting and precise experimental results presented there with those summarized in Fig.~3 of our present paper, one may see that the proposed $T^*$ also agrees with the temperature onset of the rounding effects in the measured magnetization above $T_c$. In fact, as the excellent measurements summarized in Ref.~\cite{taillefer} were performed in detwinned crystals, it could be useful to note here that the rounding behavior observed in earlier resistivity measurements, performed also in a detwinned OPT Y-123 crystal, were explained at a quantitative level in terms of the GGL approach for conventional superconducting fluctuations.\cite{ref9}  By introducing a total energy cutoff, in our present paper we have extended such an agreement up to the own temperature onset of the measured roundings.

Let us further stress that the comments on the $T^*$ location presented above just concerns the OPT Y-123 superconductor. What origin could have these discrepancies? Contrarily to the entangled difficulties that, as we have briefly commented in the Introduction of our present paper, may arise when analyzing the rounding effects on the in-plane resistivity above $T_c$ in overdoped or underdoped compounds and also in other observables, in this case the answer seems quite direct: The main argument proposed in {Ref.~\cite{taillefer}} (and also by other authors; see, e.g., Ref.~\cite{ref30a} and our comments therein) to discard the conventional superconducting fluctuations as the physical origin of the rounding effects they measure above $T_c$, in particular in the resistivity, is based on the {questionable} hypothesis that these conventional fluctuations may be quenched by relatively low reduced magnetic fields, $h = H/H_{c2}(0)$.\footnote{An illustrative example may be seen in {Ref.~\cite{taillefer}}, where such an hypothesis was explicitly used to separate between two different contributions to the negative Nernst coefficient measured above $T_c$: One contribution, only observable just near $T_c$ and ``strongly field-dependent'', which these authors attribute to superconducting fluctuations;  and a second contribution, magnetic field-independent, which extends up to the onset of the negative Nernst effect and attributed to quasiparticles in the pseudogap phase.}  However, such a criterion is questionable because the field amplitudes used in these measurements are not important enough as to quench the conventional superconducting fluctuations above $T_c$. For instance, in the case of the OPT Y-123 even the highest field amplitudes used correspond to reduced fields well below 0.1 which, as one may easily check by using any version of  the GGL approach,\cite{ref2} is too small to efficiently diminish the Gaussian fluctuations of the superconducting order parameter. In fact, one could arrive to this conclusion by just taking a look to direct experimental results obtained under much larger reduced fields in low-$T_c$ and high-$T_c$ superconductors (see our comment to Ref.~\cite{ref30a} and the corresponding references we suggest therein). 

\subsection{On the seeming discrepancies with the thermomagnetic effects around $T_c$}

As the onset of the negative Nernst coefficient measured above $T_c$ in cuprate superconductors was related to a pseudogap by different research groups (see, e.g., Ref.~\cite{taillefer}), and taking also into account the importance, stressed above, of the precise localization of $T^*$ in the OPT Y-123 superconductor, it could also be useful to briefly comment here on this aspect. Note first that the own nature of this anomalous Nernst effect is still at present a debated issue.\cite{ref1a,Ussishkin02,ref41,ref30f,ref64c} In fact, the discrepancies manifest already in the procedures proposed to separate the superconducting effects from the normal state (background) behavior: Well above $T_c$, some authors attribute the decrease of the Nernst coefficient, often observed in different cuprate superconductors, to the normal state behavior. Then, the superconducting effects are extracted by using a baseline background that eliminates that decrease. This procedure was used by different authors,\cite{Rullier06,Xu00} and it is illustrated, for instance, in Fig.~3 of the early paper of Xu and coworkers.\cite{Xu00} In contrast, other authors obtain the background of the Nernst coefficient by extrapolating up to $T_c$ its normal state behavior at high reduced temperatures. Such a procedure may be appreciated in {Fig.~1 of the first paper of Ref.~\cite{taillefer}}, where the corresponding Nernst effect was attributed to quasiparticles in the pseudogap region. 

The discrepancies and shortcomings when analyzing the thermomagnetic effects above $T_c$ in cuprate superconductors are to some extent similar to those already encountered, many years ago, in the case of the low-$T_c$ superconductors (see, e.g., Refs.~\cite{Solomon67,Fiory67,Huebener69,Vidal73}): There was a large consensus in attributing the origin of the thermomagnetic effects observed below $T_c$ to the movement of magnetic vortices and to its corresponding entropy flux. This consensus extends also to a similar origin of these effects, also below $T_c$, in the cuprate superconductors, as already stressed when analyzing earlier measurements in the cuprates (see, e.g., Refs.~\cite{Palstra90,ref27}, and references therein). In contrast, also in low-$T_c$ superconductors there was not consensus on the origin of the thermomagnetic effects observed above $T_c$, the mechanisms proposed being as different as surface superconductivity, conventional superconducting fluctuations or normal state effects.\cite{Solomon67,Fiory67,Huebener69,Vidal73} These discrepancies remain until now, but they do not have prevented a large consensus when attributing to conventional superconducting fluctuations the measured resistivity and magnetization rounding effects.\cite{ref2} A particularly illustrative example of this situation is provided by the results obtained in Pb-In alloys: Early detailed measurements of the Ettingshausen effect show an appreciable signal up to 1.25$H_{c2}(0)$, whose origin is not yet well settled \cite{Vidal73}. In contrast, the rounding effects on the magnetization measured in these alloys extend only up 1.1$H_{c2}(0)$, this last result being in excellent agreement with the GGL approach under a total energy cutoff for the conventional superconducting fluctuations.\cite{ref76,Mos01}

\section{Conclusions}

The rounding effects on the electrical conductivity and on the magnetization have been measured on both sides of the critical transition temperature, $T_c$, in {a high-quality single-crystal and in a thin film} of the prototypical optimally-doped YBa$_2$Cu$_3$O$_{7-\delta}$ high-$T_c$ superconductor, including the high reduced-temperature region above $T_c$ and the low and moderate magnetic field regimes. An aspect particularly remarkable of our present measurements is that the high chemical and structural quality of our samples has considerably reduced the ambiguities associated with the corresponding $T_c$-inhomogeneities. In addition, the comparison between both observables provides an useful consistency check. The direct experimental information, model and background independent, provided by our measurements includes the temperature location of the rounding onset above $T_c$, that is found to be $T_{\rm onset}/T_c=1.7\pm 0.2$ for all the studied observables. When analyzing in the fourth part of our paper these measurements in terms of the GL scenario {with} superconducting fluctuations, the comparisons between the roundings of the resistivity and of the magnetization in the zero field limit provide an experimental demonstration of the absence of indirect fluctuation effects  on the paraconductivity (in particular, Maki-Thomson or DOS effects), and of non-local electrodynamic effects on the precursor diamagnetism, even in the high reduced-temperature region.

The analyses {of the high quality experimental data summarized here} in terms of the GL scenario include the extended (with a total-energy cutoff) Gaussian-Ginzburg-Landau (GGL) approach for multilayered superconductors above $T_c$ and, consistently, the GL-lowest-Landau-level (GL-LLL) scaling for 3D superconductors in the critical region around $T_c(H)$. Our results confirm at a quantitative level the adequacy of these phenomenological descriptions, and they {strongly suggest} that the possible presence of other, unconventional, mechanisms do not appreciably affect the rounding of the different observables above $T_c$ in OPT Y-123. 
{These analyses contribute to clarify two central aspects, still open and debated until now, of the GGL scenario:  They support at a quantitative level that the effective interlayer distance in this multilayered superconductor may be approximated as one half of the crystallographic periodicity length, as proposed in Ref.~\cite{ref5} by studying at a theoretical level the GGL approach for multilayered superconductors; and they show the adequacy of the so-called total energy cutoff, that takes into account the Heisenberg quantum localization energy associated with the shrinkage of the superconducting wave function when the temperature increases well above $T_c$. This total-energy cutoff not only accounts for the observed rounding onset temperatures but also explains at a quantitative level the behavior of these roundings below their onsets, in the so-called high reduced-temperature region for the superconducting fluctuations.}

{Complementarily, the comparisons between our present experimental results and the GL approach provide precise information on the parameters arising in that scenario, including the} in-plane and perpendicular superconducting coherence length amplitudes, or the relative (to the BCS prediction) relaxation time of the fluctuating superconducting pairs, $\tau _{\rm rel}$. Our present results also further confirm that both $T_{\rm onset}/T_c$ and $\tau _{\rm rel}$ arising in this extended GGL scenario take values similar to those that have been measured in conventional low-$T_c$ superconductors,\cite{ref76} which in turn are similar to those that may be estimated from the conventional BCS approach.

Finally, we have also briefly commented on some existing proposals to explain these rounding effects in terms of two unconventional (non-GL) scenarios: the charge density waves and the presence of a pseudogap. We conclude that, in the case of OPT Y-123, at present the experiments do not unambiguously support that these effects are the origin of the rounding above $T_c$. On the contrary, the results summarized in our present paper, in particular in the high reduced-temperature region, support that the measured roundings are originated by the presence of conventional fluctuations of the superconducting order parameter, well described by the Gaussian-Ginzburg-Landau approaches. As we have stressed above, these conclusions directly concern some of the proposals for the pairing mechanisms in cuprates. Moreover, the interest of our present results, in particular the extensions of the theoretical GGL scenario to high reduced temperatures, is also enhanced by their usefulness in future analyses at a quantitative level of the rounding effects around $T_c$ in underdoped and overdoped samples. However, as also stressed above, the unavoidable presence of chemical inhomogeneities \cite{ref17, ref30k, ref36}, that lead to important inhomogeneities of $T_c$, and also the possible entanglement with the effects associated with the opening of a pseudogap well above the measured $T_c$, will add further difficulties when performing these analyses. {In turn, these last conclusions further enhance} the relevance of the prototypical OPT Y-123 to probe the phenomenological descriptions of the superconducting transition in cuprates.

\section{Acknowledgments}

This work was supported by project FIS2016-79109-P (AEI/FEDER, UE), and the Xunta de Galicia (projects GPC2014/038 and AGRUP 2015/11). JCV acknowledges financial support from the Spanish Ministry of Education (grant FPU14/00838). We acknowledge S. Blanco-Canosa for useful conversations about CDW in OPT Y-123.

\appendix

\section{Extension to high reduced-temperatures of the GGL approach for the fluctuation conductivity and the precursor diamagnetism in 2D and layered superconductors}

\subsection{Fluctuation conductivity}

Recently\cite{ref87} the fluctuation-induced conductivity and magnetoconductivity have been calculated in the finite magnetic field regime in 3D superconductors by combining the standard GL expression for the thermal-averaged current density of the superconducting condensate with the generalized Langevin equation of the order parameter, taking into account a total-energy cutoff to include the short-wavelength effects expected to be important when $\varepsilon\stackrel{>}{_\sim}0.1$.\cite{ref87} This procedure is similar to the one earlier proposed by A. Schmid in his pioneering calculations in the zero field limit and without cutoff.\cite{ref71} Here, we briefly summarize a similar extension of these calculations to the 2D and the layered scenarios.

We first consider the 2D case, {\it i.e.}, a superconductor with thickness $d$ in the perpendicular (or $z$-) direction much smaller than the superconducting coherence length amplitude in that direction [{\it i.e.}, $d \ll \xi_c(0)$]. Under this condition the  spectrum of the fluctuations appearing in the formalism of Ref.~\cite{ref87}  becomes frustrated along the $z$-direction (with the corresponding component of the fluctuations wavevector, $k_z$, bound between $-\pi/d$ and $\pi/d$ and verifying $\xi_c(0) k_{z}\ll1$). When applied to Eqs.~(B.17) and (B.18) of Ref.~\cite{ref87}, this leads to
\begin{equation}
\Delta \sigma_{2D}(\varepsilon,h,\varepsilon^c)=
\frac{e^2   }{32 \hbar d}
\frac{1}{h}
\left[
\psi^{1}\left(\frac{\varepsilon+h}{2h}
\right)-
\psi^{1}\left(\frac{\varepsilon^c+h}{2h}
\right)
\right]
\label{cutmagnetoprange2D}
\end{equation}
for the fluctuation-induced conductivity in the finite magnetic field regime with a total-energy cutoff, and to
\begin{equation}
\Delta \sigma_{2D}(\varepsilon,h)=
\frac{e^2}{32 \hbar d}
\frac{1}{h}
\psi^{(1)}\left(\frac{\varepsilon+h}{2h}\right)
\label{magnetoprangedosd}
\end{equation}
in absence of cutoff.

In the zero field limit ({\it i.e.} for $h \ll \varepsilon,\varepsilon^c$), from Eq.~\ref{cutmagnetoprange2D} it results that the paraconductivity with a total-energy cutoff in a 2D superconductor is
\begin{eqnarray}
\Delta \sigma_{2D}(\varepsilon,\varepsilon^c)=
\frac{e^2  }{16 \hbar d}\left(\frac{1}{\varepsilon}-\frac{1}{\varepsilon^c}\right)\, \, ,
\label{sigma2dener}
\end{eqnarray}
that in absence of cutoff leads to the well-known Aslamazov-Larkin (AL) expression for $\Delta \sigma$ in a 2D superconductor\cite{ref88}
\begin{eqnarray}
\Delta \sigma_{2D}(\varepsilon)=
\frac{e^2  }{16 \hbar d \varepsilon}\, \, .
\label{sigma2dal}
\end{eqnarray}

Let us note here that, as done in the classical paper by Hikami and Larkin\cite{ref89}  for the low-$\varepsilon$, no-cutoff case, we could expand Eqs.~\ref{cutmagnetoprange2D} and (\ref{magnetoprangedosd}) in powers of $h$ up to quadratic order. Our Eq.~\ref{magnetoprangedosd} would then produce for $\Delta\tilde{\tilde{\sigma}}$ an extra prefactor $2/3$ with respect to the result in Ref.~\cite{ref89}, which origin may be traced back to a different use of the Langevin  equation for the order parameter in both calculations. (This prefactor difference exist also in the layered case, presented later in this Appendix). It does not seem feasible in practice to determine what use of the Langevin equation produces  better accordance with experiments at low fields, given that the prefactor may be compensated with a change of only  $\sim10$\% in the value of $\xi_{ab}(0)$ [because $h\propto\xi_{ab}^{2}(0)$] and this is below the  uncertainty in any  determination of the experimental value of $\xi_{ab}(0)$. However, direct introduction of the cutoff in the formalism of Ref.~\cite{ref89} leads to some unphysical inconsistencies in the resulting equations; we favor, therefore, our present formalism when analyzing the data including the $\varepsilon\stackrel{>}{_\sim}0.1$ region.

We now summarize here the extension of these results to  layered superconductors.\cite{CaLaFeNiAs} We shall consider here only layered superconductors with  a single interlayer distance, $s$. For that, we  first adapt Eq.~(B.18) of Ref.~\cite{ref87} (giving the fluctuation-induced conductivity in 3D as a sum over the contributions of  different Landau levels) to the layered case by introducing the appropriate  out-of-plane spectrum of the fluctuations\cite{ref89} ({\it i.e.}, substituting $\omega_{k_{z}}^{3D}=\xi_c^2(0)k_{z}^2$ by $\omega_{k_{z}}^{\rm LD}=B_{LD}[1-\cos(k_{z}s)]/2$) and taking into account the structural cutoff in the $z$-direction through $|k_z|\leq \pi/s$. This leads to
\begin{equation}
\Delta \sigma_{LD}=
\frac{e^2h}{16 \pi \hbar }
\int_{-\pi/s}^{\pi/s}{\rm d}k_{z}
\sum_n [\varepsilon+h(2n+1)+\omega_{k_{z}}^{LD}]^{-2},
\label{summagnetoLD}
\end{equation}
where  the sum over Landau-levels is to be performed up to $n_{max}=(\varepsilon^c-\varepsilon)/(2h)-1$.\setcounter{footnote}{0}\footnote{Previous calculations of different fluctuation-induced observables under a total energy cutoff in layered superconductors were obtained from the corresponding $2D$ results by applying the substitution $\varepsilon\to\varepsilon+\omega_{k_z}^{LD}$ even to the maximum Landau level index, that leads to $n_{max}=\frac{\varepsilon^c-\varepsilon-\omega_{k_{z}}}{2h}-1$.\cite{ref18,ref19,ref20} In the 3D limit $\omega_{k_z}^{LD}\to \omega_{k_z}^{3D}$ and the condition on $n_{max}$ is somehow equivalent to cut the $k$-space considering only the fluctuating modes inside a sphere. However, to be consistent with our own 3D calculations of $\Delta\sigma(\varepsilon,h)$ and of other fluctuation-induced observables (see Ref.~\cite{ref87} and references therein), we have performed the total energy cutoff in the form of a finite cylinder, that leads to $n_{max}=\frac{\varepsilon^c-\varepsilon}{2h}-1$. The difference between the two ways to apply the total energy cutoff in the LD-scenario lead to small quantitative differences that can be absorbed by a small change in the cutoff amplitude in materials where $B_{LD} \ll \varepsilon^c$, as it is the case of OPT Y-123.}
Then, we obtain
\begin{equation}
\hspace{-2cm}\Delta \sigma_{LD}(\varepsilon,h,\varepsilon^c)=\frac{e^2}{64 \pi \hbar h}\int_{-\pi/s}^{\pi/s}{\rm d}k_{z}
\left[\psi^{1}\left(\frac{\varepsilon+h+\omega_{k_{z}}^{LD}}{2h}\right) 
-\psi^{1}\left(\frac{\varepsilon^c+h+\omega_{k_{z}}^{LD}}{2h}\right)
\right]
\label{cutmagnetoprangeLD}
\end{equation}
for the fluctuation-induced conductivity in layered superconductors under an applied magnetic field with a total-energy cutoff. The paraconductivity in the same
scenario corresponds to the limit $h \ll \varepsilon,\varepsilon^c$ of the above expression, given by
\begin{equation}
\Delta \sigma_{LD}(\varepsilon,\varepsilon^c)=\frac{e^2  }{16 \hbar s}
\left[\frac{1}{\varepsilon}\left(1+\frac{B_{LD}}{\varepsilon}\right)^{-1/2}
-\frac{1}{\varepsilon^c}\left(1+\frac{B_{LD}}{\varepsilon^c}\right)^{-1/2}\right]\,.
\label{sigmaLDener}
\end{equation}
In 2D and 3D superconductors, in the long-wavelength regime ({\it i.e.}, for $h,\varepsilon\ll \varepsilon^c$) Eqs.~\ref{cutmagnetoprangeLD} and (\ref{sigmaLDener}) become cutoff
independent and they reduce to
\begin{equation}
\Delta \sigma_{LD}(\varepsilon,h)= 
\frac{e^2}{64 \pi \hbar} \frac{1}{h} \int_{-\pi/s}^{\pi/s}{\rm d}k_{z}\psi^{1}\left(\frac{\varepsilon+h+\omega_{k_{z}}^{LD}}{2h}\right),
\label{magnetoprangeLD}
\end{equation}
and, respectively,
\begin{equation}
\Delta \sigma_{LD}(\varepsilon)=
\frac{e^2  }{16 \hbar s}\frac{1}{\varepsilon}\left(1+\frac{B_{LD}}{\varepsilon}\right)^{-1/2}\, \, .
\label{sigmaLD}
\end{equation}
This last being the well-known AL-like expression for the paraconductivity in the LD-scenario.\cite{ref79}

As already commented in detail for the 3D case in Ref.~\cite{ref87}, the high-$\varepsilon$ behaviour of the zero-field paraconductivity produced by our present formalism is  slightly different from our previous results for $\Delta\sigma_{ab}(\varepsilon,\varepsilon^c)$ in Ref.~\cite{ref18}. These small differences are only quantitative, not qualitative: In both cases it is predicted a rapid falloff of the fluctuation-induced effects above $\varepsilon \simeq 0.1$, and a vanishing of these fluctuations at $\varepsilon=\varepsilon^c$. In fact, in our present analysis these differences could be absorbed by a small change in the $\varepsilon^c$ value. 

\subsection{Fluctuation-induced diamagnetism}

Let us here summarize the $\Delta M$ expressions calculated as described in Ref.~\cite{ref19} but under the cutoff prescriptions described in the previous subsection. We obtain for layered and 2D superconductors respectively:
\begin{eqnarray}
\Delta M_{LD}(\varepsilon,H)&=&\frac{-k_B T}{2 \pi \phi_0} \int_{-\pi/s}^{\pi/s} \left[-\frac{\varepsilon^c+w^{\rm LD}_{k_z}}{2h} \psi\left(\frac{h+\varepsilon^c+w^{\rm LD}_{k_z}}{2h}\right)\right.\nonumber\\ 
&&\left.-\ln \Gamma\left(\frac{h+\varepsilon+w^{\rm LD}_{k_z}}{2h}\right)+\ln\Gamma \left(\frac{h+\varepsilon^c+w^{\rm LD}_{k_z}}{2 h}\right)\right.\nonumber\\ 
&&\left.+\frac{\varepsilon+w^{\rm LD}_{k_z}}{2h}\psi\left(\frac{h+\varepsilon+w^{\rm LD}_{k_z}}{2h}\right)+\frac{\varepsilon^c-\varepsilon}{2h}\right]{\rm d}k_z,
\end{eqnarray}
and 
\begin{eqnarray}
\Delta M_{2D}(\varepsilon,H)&=&\frac{-k_B T}{\phi_0 s} \left[-\frac{\varepsilon^c}{2h} \psi\left(\frac{h+\varepsilon^c}{2h}\right)-\ln \Gamma\left(\frac{h+\varepsilon}{2h}\right)\right.\nonumber\\ 
&&\left.+\ln\Gamma \left(\frac{h+\varepsilon^c}{2 h}\right)+\frac{\varepsilon}{2h}\psi\left(\frac{h+\varepsilon}{2h}\right)+\frac{\varepsilon^c-\varepsilon}{2h}\right].  
\end{eqnarray}

By applying $h \ll \varepsilon,\varepsilon^c$ in the above equations, we obtain the Schmidt limit under a total-energy cutoff
\begin{equation}
\Delta M_{LD}(\varepsilon)=\frac{-k_B T}{6 \phi_0 s} h \left[\frac{1}{\varepsilon}\left(1+\frac{B_{LD}}{\varepsilon}\right)^{-1/2}-\frac{1}{\varepsilon^c}\left(1+\frac{B_{LD}}{\varepsilon^c}\right)^{-1/2}\right],
\label{eqB17}
\end{equation}
and
\begin{equation}
\Delta M_{2D}(\varepsilon)=\frac{-k_B T}{6 \phi_0 s} h\left(\frac{1}{\varepsilon}-\frac{1}{\varepsilon^c}\right).
\label{eqB18}
\end{equation}

In the limit $\varepsilon\ll\varepsilon^c$ these expressions recover the classical expressions for the Schmidt limit for layered and 2D superconductors respectively:\cite{ref5,ref6,ref19,ref78,ref79,ref80,ref81}
\begin{eqnarray}
\Delta M_{LD}(\varepsilon) = -\frac{k_B T}{6 \phi_0 s}\frac{h}{\varepsilon}\left(1+\frac{B_{LD}}{\varepsilon}\right)^{-1/2},
\end{eqnarray}
and 
\begin{eqnarray}
\Delta M_{2D}(\varepsilon) = -\frac{k_B T}{6 \phi_0 s}\frac{h}{\varepsilon}.
\end{eqnarray}

\end{document}